\documentclass[twocolappendix,numberedappendix,iop]{emulateapj}
\usepackage{graphicx}
\pdfoutput=1

\usepackage[backref,breaklinks,colorlinks=true,citecolor=blue]{hyperref}
\usepackage[all]{hypcap}

\usepackage{etoolbox}

\makeatletter

\patchcmd{\NAT@citex}
  {\@citea\NAT@hyper@{%
     \NAT@nmfmt{\NAT@nm}%
     \hyper@natlinkbreak{\NAT@aysep\NAT@spacechar}{\@citeb\@extra@b@citeb}%
     \NAT@date}}
  {\@citea\NAT@nmfmt{\NAT@nm}%
   \NAT@aysep\NAT@spacechar\NAT@hyper@{\NAT@date}}{}{}

\patchcmd{\NAT@citex}
  {\@citea\NAT@hyper@{%
     \NAT@nmfmt{\NAT@nm}%
     \hyper@natlinkbreak{\NAT@spacechar\NAT@@open\if*#1*\else#1\NAT@spacechar\fi}%
       {\@citeb\@extra@b@citeb}%
     \NAT@date}}
  {\@citea\NAT@nmfmt{\NAT@nm}%
   \NAT@spacechar\NAT@@open\if*#1*\else#1\NAT@spacechar\fi\NAT@hyper@{\NAT@date}}
  {}{}

\makeatother

\begin{document}
\title{Dust and Gas in the disk of HL Tauri:\\ Surface density, dust settling, and dust-to-gas ratio}

\author{C.~Pinte\altaffilmark{1,2}}
\email{christophe.pinte@obs.ujf-grenoble.fr}
\author{W.R.F.~Dent\altaffilmark{3}}
\author{F.~M\'enard\altaffilmark{1,2}}
\author{A.~Hales\altaffilmark{3,4}}
\author{T.~Hill\altaffilmark{3}}
\author{P.~Cortes\altaffilmark{3,4}}
\author{I.~de Gregorio-Monsalvo\altaffilmark{3}}

\altaffiltext{1}{UMI-FCA, CNRS/INSU, France (UMI 3386), and Dept. de
  Astronom\'{\i}a, Universidad de Chile, Santiago, Chile.}
\altaffiltext{2}{Univ. Grenoble Alpes, IPAG, F-38000 Grenoble, France
CNRS, IPAG, F-38000 Grenoble, France}
\altaffiltext{3}{Atacama Large Millimeter / Submillimeter Array, Joint ALMA Observatory, Alonso de C\'ordova 3107, Vitacura 763-0355, Santiago, Chile}
\altaffiltext{4}{National Radio Astronomy Observatory, 520 Edgemont Road, Charlottesville, Virginia, 22903-2475, United States}

\date{}

\begin{abstract}
The recent ALMA observations of the disk surrounding HL~Tau reveal a
very complex dust spatial distribution.
We present a radiative transfer model
accounting for the
observed gaps and bright rings as well as radial changes of the
emissivity index. We find that the dust density is
depleted by at least a factor of 10 in the main gaps compared to the surrounding
rings. Ring masses range from 10--100\,M$_{\oplus}$ in dust, and
we find that each of the deepest gaps
is consistent with the removal of up to 40\,M$_{\oplus}$
of dust. If this material has accumulated into rocky bodies, these would
be close to the point of runaway gas accretion.
Our model indicates that
the outermost ring is
depleted in
millimeter grains compared to the central rings.
This suggests faster
grain growth in the central regions and/or radial migration of the larger
grains.
The morphology  of the gaps observed by ALMA - well separated and showing a high degree of contrast with the
bright rings over all azimuths -
indicates that the millimeter dust disk is
geometrically thin (scale height $\approx$\,1\,AU
at 100\,AU) and that a large amount of settling of large grains has already
occurred.
Assuming a standard dust settling model, we find that the
observations are consistent with a turbulent viscosity coefficient of a
few $10^{-4}$.
We estimate the gas/dust ratio in this thin layer
to be of the order of 5 if the initial ratio is 100.
The HCO$^+$ and CO emission is consistent with gas in
Keplerian motion around a 1.7\,$M_\odot$ star at radii from $\leq10 - 120\,$AU.
\end{abstract}
\maketitle

\defcitealias{ALMA_HLTau}{ALMA15}


\section{Introduction}
Planet formation within disks around young stars requires the
growth of sub-micron dust grains over many orders of magnitude.
The dust feeding the disk in its early evolution is thought to have grown from
the submicron grains typical of the interstellar medium, up to micron sized particles in the dense regions of molecular clouds and cores
\citep[e.g.][]{Pagani10}.
Once incorporated into the disk, these grains must then grow from micron-sized
to pebbles and ultimately kilometer-sized bodies. This extensive
growth phase can only occur in the highest density regions of a protoplanetary
disk -- at or close to the disk's midplane
\citep{Dominik07PPV,Testi14PPVI}.

The gas drag acting on the dust as it orbits the central star causes it to
settle rapidly to the midplane. Just how much remains a poorly constrained
  quantity by direct observations.
The small dust particles ($<$ 1\,$\mu$m), best traced by scattered light, remain coupled
tightly to the gas, while the larger particles, best studied at longer
wavelengths, move toward the midplane. The thickness of the midplane
depends on the balance between vertical settling and counter-acting stirring
mechanisms \citep{Garaud2004,dd05,Fromang06}.
Once in the midplane, particles are also expected to drift radially inward
as a result of the combined action of the stellar gravity, the gas drag and the radial pressure gradient \citep{Barriere05,Birnstiel10}.
The presence of planets within the disk further complicates the dynamics of the
dust grains as they may create gaps whose profile
depends on the planet mass, density profile, and grain size \citep{Paardekooper04,Fouchet07,Gonzalez12}.

Fitting multi-color scattered light images of edge-on disks indicates the presence of stratification,
with smaller grains found higher up in the disk atmosphere while larger grains
have settled deeper \citep[e.g.][]{Pinte07,Pinte08b,Duchene10}, as predicted by models.
Tracing the properties of the
dust layers deep inside the disk requires observations at long (millimeter)
wavelengths where the disk becomes optically thin.
Several multi-wavelength studies are now suggesting that the larger grain abundances depend on radius, with
a concentration toward the center of protoplanetary disks
\citep{Ricci10,Guilloteau11,Perez12,Menu14}; however, the spatial resolution was
insufficient to resolve the vertical structure of the dust layer.
In general, the midplane region of disks remains poorly constrained by direct observations.

HL Tau is a T Tauri star of spectral energy distribution (SED) Class I--II in the Taurus molecular cloud, at a distance of 140\,pc.
The disk
has been widely studied in the (sub-)millimeter wavelength continuum \citep{Beckwith90,Mundy96,Wilner96,Lay97,Chandler00,Looney00,Greaves08,Kwon11,Stephens14}.
The highest resolution millimeter observations so far \citep{Kwon11} revealed
a 120\,AU dust disk and suggested, with comparison with the SED, that the
millimeter grains may have already settled to the midplane.
In addition to the disk structure, an orthogonal optical jet and a molecular
bipolar outflow \citep[e.g.][]{Mundt90} and an envelope \citep{Menshchikov99,Robitaille07} have been reported.

HL~Tau was recently observed with ALMA  using baselines up to $\sim$ 15\,km
at wavelengths from 0.8-3\,mm, which resulted in spatial resolutions down to 3.5\,AU \citep[]
[hereafter \citetalias{ALMA_HLTau}]{ALMA_HLTau}. These data provide the critical
angular resolution needed to study the details of the disk midplane. They indicate the presence of numerous gaps and rings.

Here we model the recent ALMA data at all three wavelengths, in order
to reproduce the main features of the HL Tau disk. We interpret the results in terms of the underlying
dust distribution and evolution. We investigate whether a single unified dust
model can reproduce both the ALMA images and SED, and
compare the dust and gas disk also observed by ALMA.

\section{Observations}

HL~Tau was observed as a science verification target within the
ALMA Long Baseline Campaign; full details of the observations are given in \citetalias{ALMA_HLTau} and \cite{ALMA_LBC}.
The band 3, 6, and 7 calibrated continuum data sets and CLEANed images were downloaded
from the ALMA Science portal\footnote{http://www.almascience.org/}, together with the
tables for self-calibration, which we applied to the data.

\begin{figure}
  \includegraphics[width=\hsize]{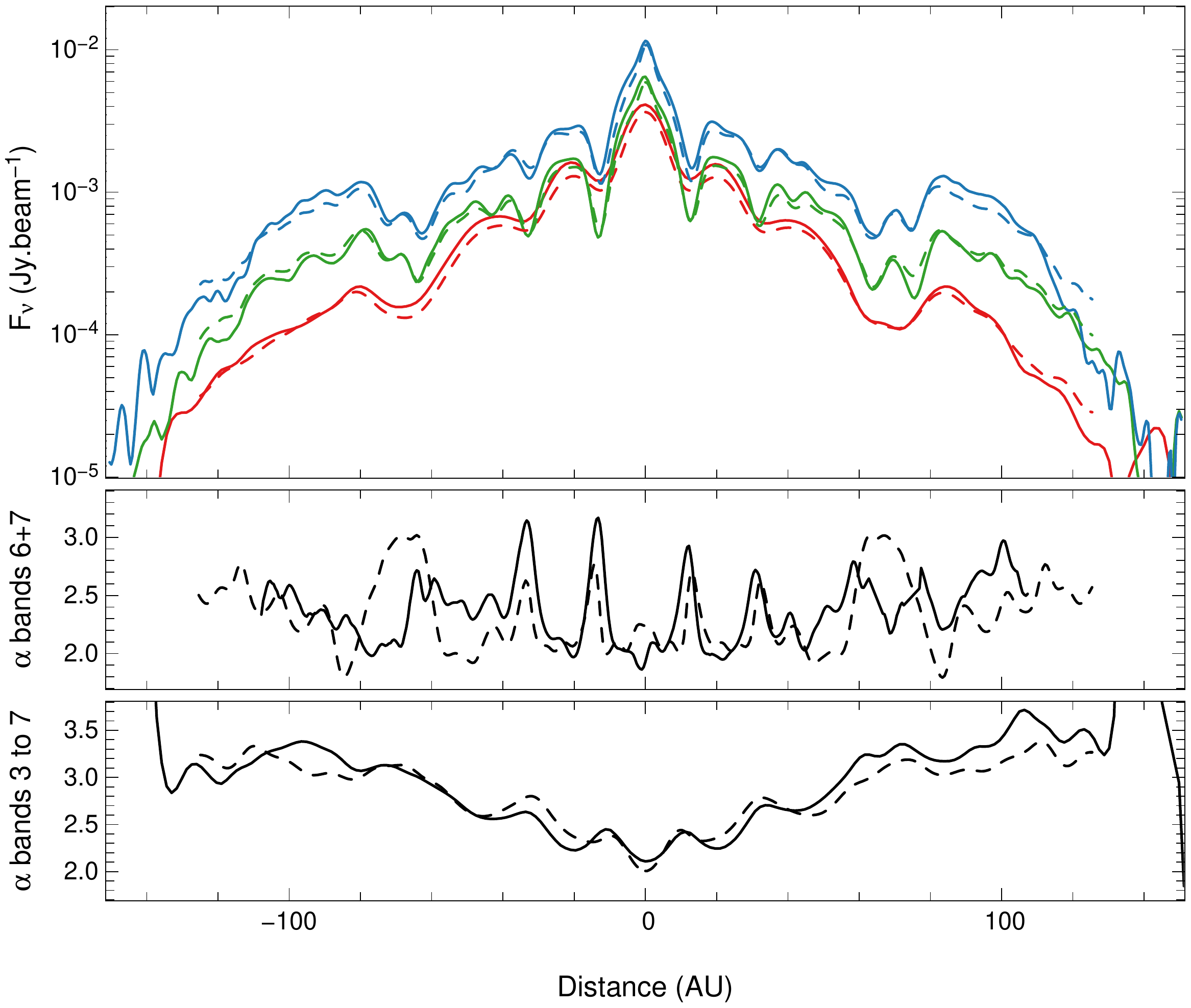}
  \caption{\emph{Top:} cuts of observed and synthetic CLEAN maps along the major axis of the disk (\emph{red} =
    band~3, 2.9\,mm, \emph{green} = band~6, 1.3\,mm, \emph{blue} = band 7,
    870$\,\mu$m.). The spatial resolution along the disk is 10, 4, and 3\,AU in
    the three bands.
    \emph{Middle:} spectral index profile derived from the combined
    6+7 bands spectral index map, from \citetalias{ALMA_HLTau}. \emph{Bottom:} Spectral index profile obtained by convolving
    the band 6 and 7 maps to the same resolution as the band 3 map and fitting
    a spectral index pixel-by-pixel. In all three panels, observations are plotted
    with a full line while the model is plotted with a dashed line.
    \label{fig:profiles}}
\end{figure}

From the images, the observed brightness in each band, and the spectral index profiles measured along the disk major axis are
presented in Figure~\ref{fig:profiles}. A similar
structure of alternating bright rings and dark gaps is seen in all three bands. The spectral index measured between 0.87 and 2.9\,mm
(lower panel) displays a significant increase from the inner regions to the
outer disk, from $\sim$2.2 to $\sim$3.5, indicating a radial
change in the spatially convolved dust emissivity and/or optical depth.
The combined cleaning of the band 6 and 7 data \citepalias{ALMA_HLTau} yields a higher resolution spectral index map, and a cut through this along the major axis is presented in the middle panel in Figure~\ref{fig:profiles}. As noted previously, this shows that the optical depths are significant in the rings at these shorter wavelengths, with sharp increases in the spectral index inside the gaps.
In order to extract the underlying dust structure, we reproduce the data at all three wavelengths, and then compare the models with these radial profiles. Any differences in the derived models would then suggest evidence of a radial change in the dust characteristics.

The calibrated spectral line data sets in band 3 for $^{12}$CO $J$=1-0 and HCO$^+$
$J$=1-0 were also downloaded and CLEANed. The default CLEAN parameters and $u$$v$
taper from
\citetalias{ALMA_HLTau} were used for HCO$^+$ giving a resolution
of 0.25\,arcsec. For the CO data set, CLEANing with a $u$$v$ taper giving a
spatial resolution of 0.2\,arcsec was found to be a good compromise between
resolving the disk and obtaining enough signal/noise per beam.
Although the SNR of the spectral line data sets is relatively low and suffers from
contamination at ambient cloud velocities, we were able to compare
the results with a simple Keplerian disk in order to investigate the gas distribution.

\section{Continuum modeling approach}

To model the ALMA continuum data, we use the radiative transfer code MCFOST
\citep{Pinte06,Pinte09}. Because of the remarkable symmetry of the ALMA images,
we assume an axisymmetric, \emph{i.e.} two-dimensional (2D), density structure. In particular, we
ignore the slight offsets and change of inclinations of the rings compared to
the central star \citepalias{ALMA_HLTau}.
We use standard disk parameters to describe the disk structure
  \citep[\emph{e.g.}][see Section~\ref{sec:RT_model}]{Pinte08,Andrews09}
except for the surface density which shape is let free and is fitted for.
Our aim is to determine whether such a simple model of the disk density structure can
reproduce the main features observed by ALMA.
Due to the complexity of the ALMA data sets, we build the model of surface
density from the images and checked a posteriori by calculating a
$\chi^2$ that the model and observed visibilities were in good agreement.

\subsection{Radiative transfer model}
\label{sec:RT_model}
We assume a flared gas density structure with a Gaussian
vertical profile $\rho_g(r,z) = \rho_0(r)\,\exp(-z^2/2\,h(r)^2)$ where
$r$ is the distance from the star in the disk midplane and $z$ is the
altitude above the midplane. For the gas disk scale
height $ h(r)$, we use a power-law distribution
$ h(r)=h_0\,(r/r_0)^{\beta}$ where $h_0$ is the scale height at
radius $r_0=100$\,AU. To keep the problem tractable, we fix some
parameters and explore the parameter space in terms of surface density and dust settling.
  The disk scale height is set to 10\,AU to be consistent with models of
thermally supported disks (\emph{e.g.} \citealp{Close97}). $\beta$ is the disk flaring exponent,
which we set to 1.15, again consistent with typical disk models. The
  exact values of $h_0$ and $\beta$  do not
strongly affect the final conclusions and the defined gas disk structure is very close to
hydrostatic equilibrium (see Appendix~\ref{app:h_hydro}). The inner
radius is set to the dust sublimation radius. The outer radius
is set to 150\,AU, \emph{i.e.} large enough that the surface density reaches
negligible values at this distance from the star and in agreement with the size
of the observed disk.

A surface density distribution per grain size $\Sigma(r,a)$ is then fitted at each
wavelength independently, using the procedure described below (Section~\ref{sec:sigma}).

Dust settling is
implemented by describing the dust transport by turbulence as a
diffusive process. This was shown to be a reasonable
approximation to global MHD simulations of stratified and turbulent disks
\citep{Fromang09}.
 The degree of dust settling is set by varying the turbulent viscosity coefficient
 $\alpha$ \citep{ShakuraSunyaev1973}. We assume that the gas
 vertical profile remains Gaussian and that the
 diffusion is constant vertically and given by $D = c_s\, h\, \alpha
 / S_c$, where $c_s = h\,  \Omega$ is the midplane sound speed, $h$ is the gas
 scale height, and $S_c$ is the
 Schmidt number, which we fix to 1.5.
 The vertical density profile for a grain of size $a$ is given by
 equation 19 of \cite{Fromang09}:

\begin{equation}
\rho(r,z,a) \propto \Sigma(r)\, \exp \left[ - \frac{\Omega \tau_S(a)}{\textit{\~{D}}} \,
\left(e^\frac{z^2}{2h^2} - 1\right) - \frac{z^2}{2h^2} \right]
\end{equation}
with $\Omega = \sqrt{\frac{GM_*}{r^3}}$, ${\textit{\~{D}}} = \frac{D}{c_s \Omega}$ and $\tau_S(a) = \frac{\rho_d a}{\rho_g c_S}$ the dust stopping time, where
$\rho_d$ is the dust material density and $\rho_g$ is the gas density in the midplane.
 The density is finally normalized so that
\begin{equation}
\int_{-\infty}^{\infty} \rho(r,z)\,\mathrm{d}z = \Sigma(r)
\end{equation}

We assume passive heating and fix the properties of the central star to
$T_\mathrm{eff} = 4\,000\,$K and $L_* = 11\,L_\odot$  \citep{Menshchikov99}.
We use a central mass $M_* = 1.7\,$M$_\odot$, as derived from the CO and HCO$^+$
  observations (see Section~\ref{sec:gas}).

We adopt a fixed dust mixture composed of 60\,\%
silicate \citep{Dorschner95}, 15\,\%
amorphous carbon \citep{Zubko96}, and 25\,\% porosity, although the details of the dust
composition do not affect the results presented in this paper. We use a grain size distribution
$\mathrm{d}n(a) \propto a^{p}\mathrm{d}a$ between $a_\mathrm{min}$
and $a_\mathrm{max}$.  $a_\mathrm{min}$ and  $a_\mathrm{max}$  are set to
$0.03\,\mu$m and 3\,mm and we initially
use $p=-3.5$ (integrated over the whole disk). However, note that $a_\mathrm{max}$ and $p$ are both
modified locally due to the dust settling and varying surface density as a
function of grain size. Each grain size is represented by a
distribution of hollow spheres with a maximum void fraction of $0.8$.

\begin{figure*}
  \includegraphics[height=0.29\hsize]{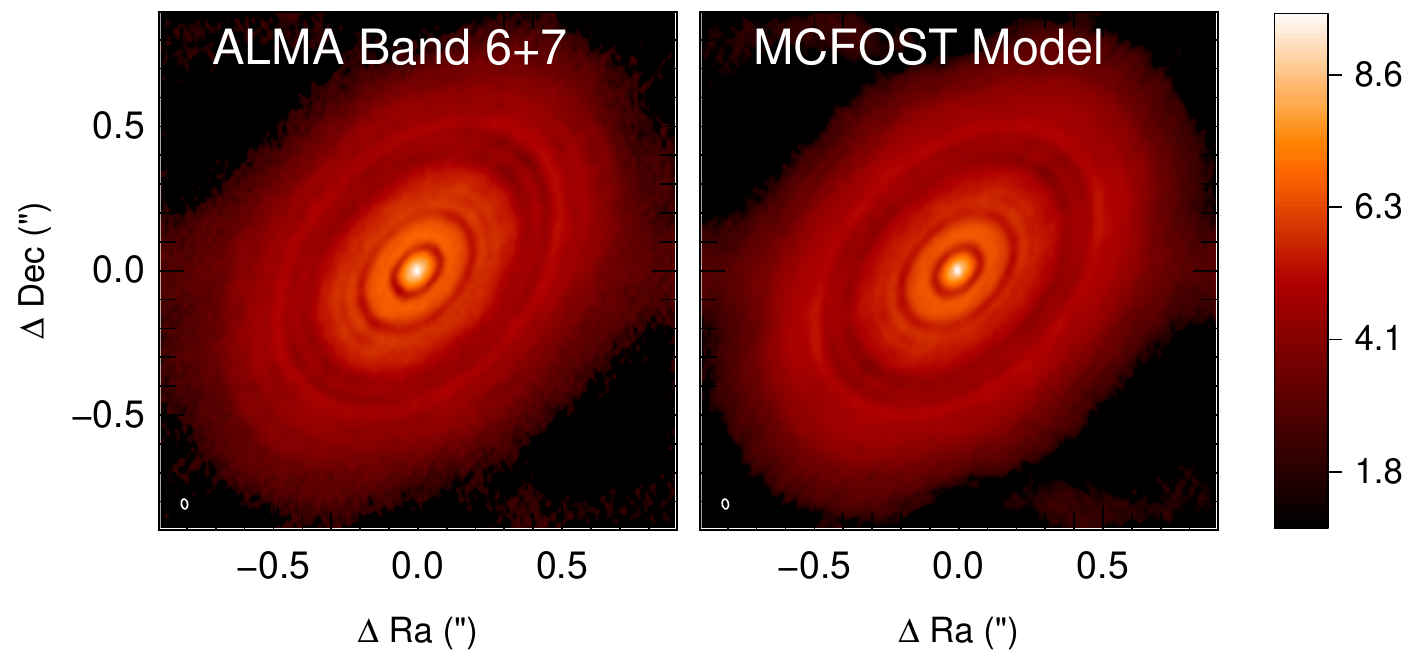}
  \hfill
  \includegraphics[height=0.29\hsize]{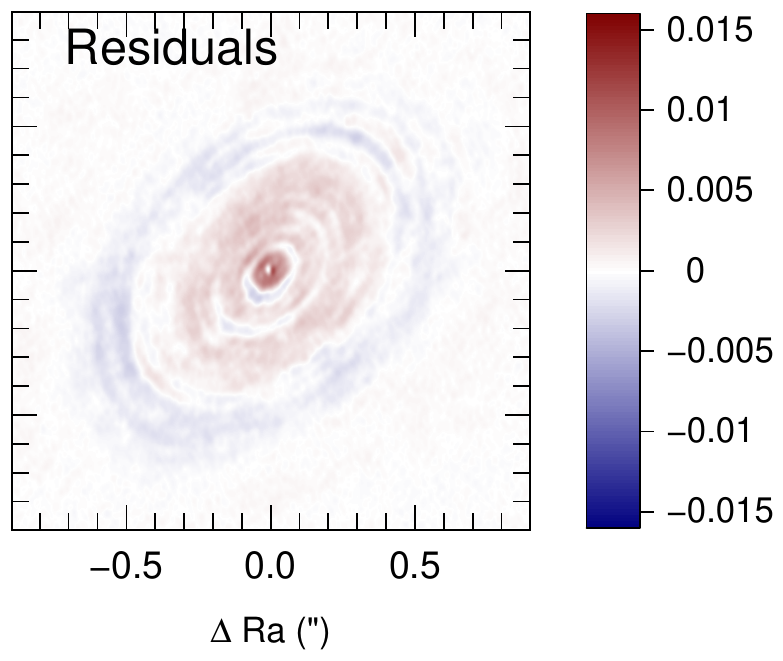}
  \includegraphics[width=0.47\hsize]{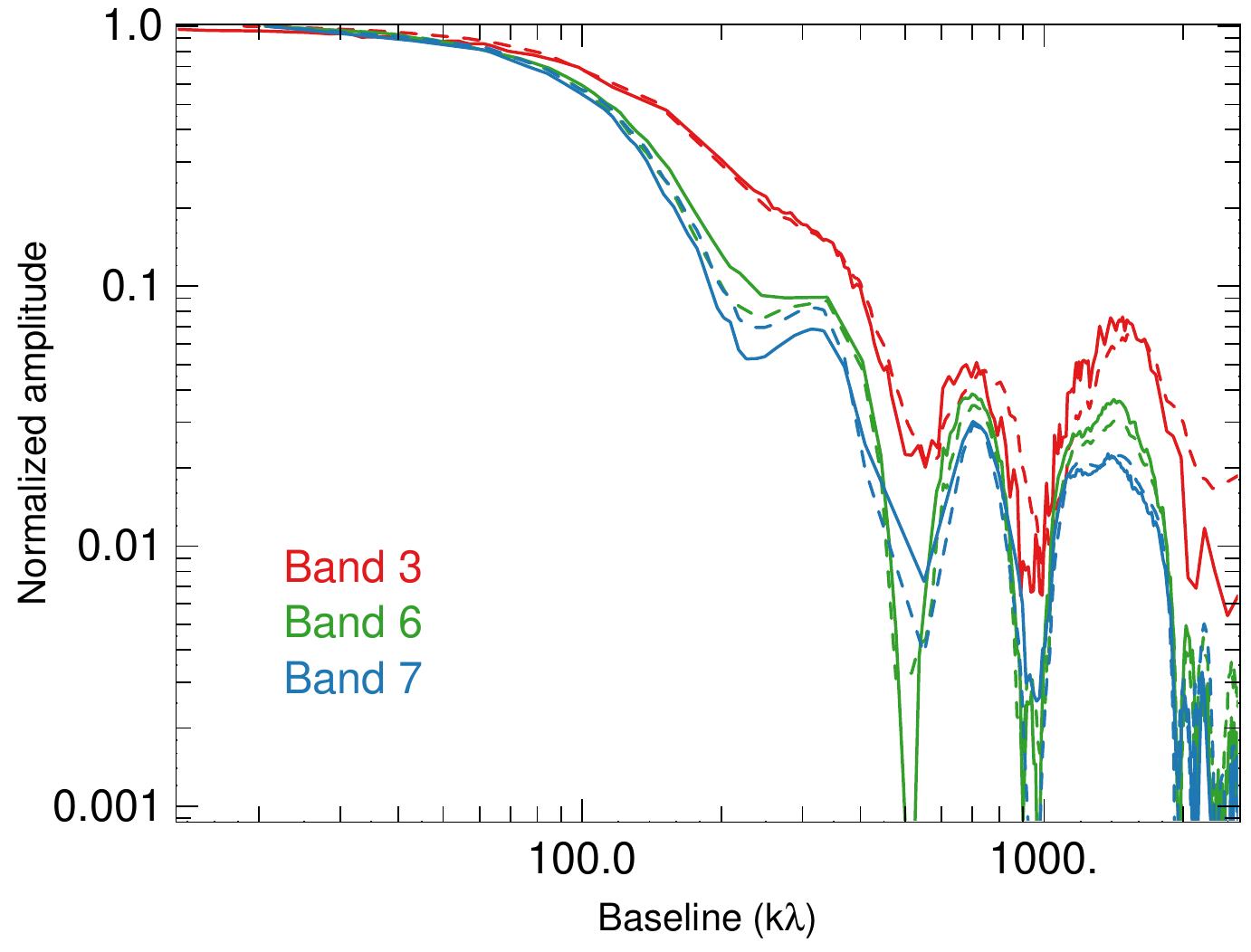}
  \hfill
  \includegraphics[width=0.47\hsize]{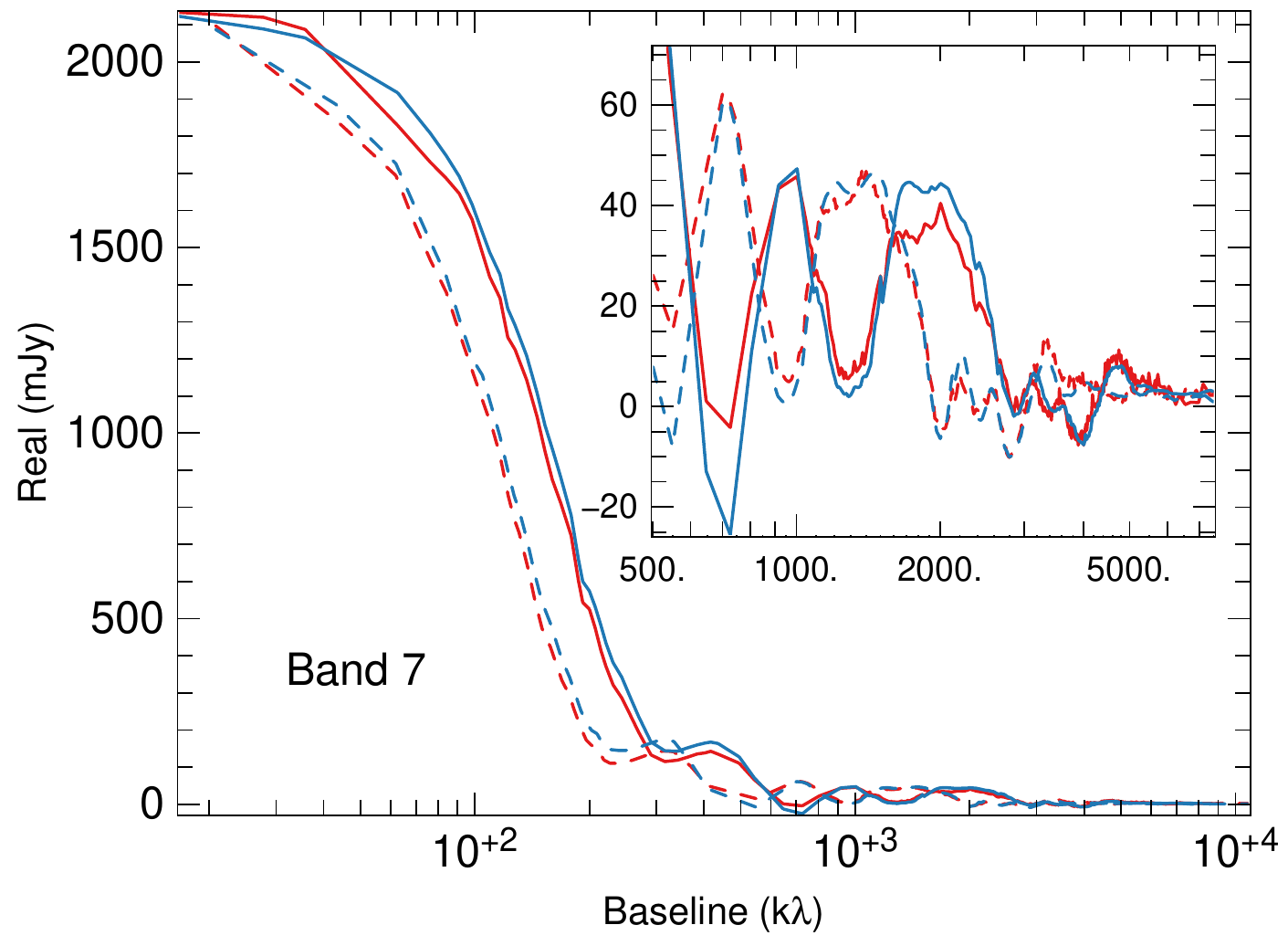}
  \caption{\emph{Upper:} comparison of the observed and model clean maps in band 6+7. The scale
  units are mJy.beam$^{-1}$. Note
    the different scale for the residual panel, which ranges from $-1\sigma$ to
    $+1\sigma$, where $\sigma$ is the standard deviation in the combined image.
    \emph{Lower left:} comparison between the observed (full lines) and
      model (dashed lines) visibility amplitudes along the disk major axis
      in the first spectral window of each
      band (band~3: red, band~6: green, band~7: blue). The model
    visibilities have been obtained from the MS
    file generated by the {\sc CASA} simulator using the exact same antenna positions
    as in the observed data. The bins were constructed to all have the same
    number of visibility points. All the uv data points within $30^\circ$ of the
    position of the major axis where considered.
    \emph{Lower right:} comparison between the observed (red) and model (blue)
    real parts of the visibilities in the first spectral window of band 7
    (337.494\,GHz; other spectral windows give similar results). The inset is a
    zoom on the baselines larger than 500\,k$\lambda$. The full lines correspond to the disk minor
    axis while the dashed lines correspond to the major axis (uv points within
    30$^\circ$ of each axis were considered). \label{fig:obs_vs_mod}}
\end{figure*}

In summary, we fixed most of the model parameters, and focused our
analysis on the parameters that are mainly probed by the new ALMA data:
the disk surface density per grain size $\Sigma(r,a)$ and dust
settling via the $\alpha$ parameter.

\subsection{Building a model of the HL Tau disk}

The construction of the 2D disk density structure was performed in two
successive steps. First, we build the surface density of the model by
reproducing the
brightness distribution along the major axis, where
the line-of-sight confusion due to the projection of the vertical structure is
minimal. This procedure is repeated for the three ALMA bands and the resulting
surface densities, which correspond to various grain sizes, are combined in a
single model.
In a second step, the vertical distribution is explored by modeling the
complete 2D data sets and quantitatively comparing the synthetic and observed visibilities.
The final model shows residuals smaller than the map rms, and visibilities in very good agreement,
indicating that the model resulting from this procedure can account for most
of the observed features in the ALMA data sets.

To obtain a surface density profile that matches the observed images, we use an
iterative procedure.
We first extract a brightness profile along the major axis
of the disk in the CLEANed map and take an average of both sides. Because the model will be convolved before
comparison with observations, we deconvolve the extracted brightness profile by
a Gaussian beam whose FWHM
corresponds to the projected FWHM along the disk semi major-axis of the
observed 2D beam. This
deconvolved brightness profile is then converted to a surface density profile
assuming an initial power-law radial temperature profile (exponent = -0.5)
consistent with the radial temperature dependence in typical disk models
\citep{DAlessio98,Dullemond02,Pinte14}. The assumption of power-law temperature
with exponent -0.5 is needed only for to the first iteration,  $T(r,z)$ is
calculated for all subsequent iterations.
Using this
initial surface density profile, we compute a model using MCFOST, convolve it with the
corresponding beam, and extract a model brightness profile along the major axis. The
surface density at any point in the disk is then corrected by the ratio of the
observed profile divided by the model profile. To avoid too many
oscillations in this procedure, this ratio is limited between 0.8 and 1.2. The
procedure is iterated until the model brightness profile does not change by
more than 1\,\% at any radius. The procedure requires of the order of 30 iterations to
reach convergence.

Note that the procedure breaks down in the very inner regions of
the disk due to high optical depth and because the central beam encloses all
the azimuthal angles. For the first five central AU, we fixed instead a power-law surface
density profile.  We find that a slope of -0.5 reproduces well the
brightness distribution of the outer parts of the central peak, but the exact value is not well
constrained because the region is only partially resolved.

Despite exploring a range of parameter space for both the disk geometry and the
dust properties, we could not reproduce the compact and peaked
emission at the position of the star.
Because the emission in the central peak is optically thick, we conclude that an additional unresolved source
such as coronal free-free emission or extra heating due to accretion might be
present close to the the star.
We added the required unresolved emission at the position of the star in the model maps: 1.57, 1.90,
and 2.59\,mJy at band 3, 6 and 7 respectively. This represents 2\,\%, 0.2\,\% and 0.1\,$\%$
of the total flux in bands 3, 6, and 7 respectively.
Note that the relative contribution of this additional emission in the
central beam is not well constrained and depends on the model parameters:
the non-resolved contributions cannot be large, but their exact value should be taken with caution.

For the iterative procedure, model images were produced by convolving
the MCFOST images by a Gaussian beam of equivalent FWHM.
The final simulated visibilities and
images were produced by applying a modified version of the {\sc CASA} simulator to
the model. This
computes the model visibilities at the exact same ($u$,$v$) coordinates as
the observations using the \emph{ft} task. To compute the model visibilities,
  we first average the observed data in time (using 300s bins)
and  in
frequency to obtain one single continuum channel per available
spectral window. This was performed to reduce
the number of ($u$,$v$) points and corresponding computing time of the {\sc
  CASA} simulator.
The thermal noise at each wavelength
was set using the median value of the precipitable water vapor (PWV) of the
observations: 1.3, 0.65, and 0.54\,mm at bands 3, 6, and 7 respectively. The resulting data set of model visibilities
was CLEANed using the same
parameters as for the observations, and the observed and modeled images were
compared for consistency (Figure~\ref{fig:obs_vs_mod}). We
estimate the quality of the model by calculating a $\chi^2$ between the
observed and synthetic visibilities. After binning, we have 229\,910, 420\,668
and  571\,960 visibilities in bands 3, 6, and 7 respectively. $\chi^2$ were
computed using the formal uncertainties from the interferometric weights, to which
we added quadratically a 10\,\% calibration uncertainty. The $\chi^2$ from the three
individual bands bands were added to compute the final $\chi2$. Our final model
has a total $\chi^2$ of 1282.
For the final model, visibilities
were also computed without performing any time averaging, and we checked that the
reconstructed simulated images almost did not show any difference, and that the
$\chi^2$ were not affected by more than 2\,\%.

\section{Results from modeling}

\subsection{Surface density profile and radial concentration of large grains}
\label{sec:sigma}

\begin{figure*}
    \includegraphics[width=\hsize]{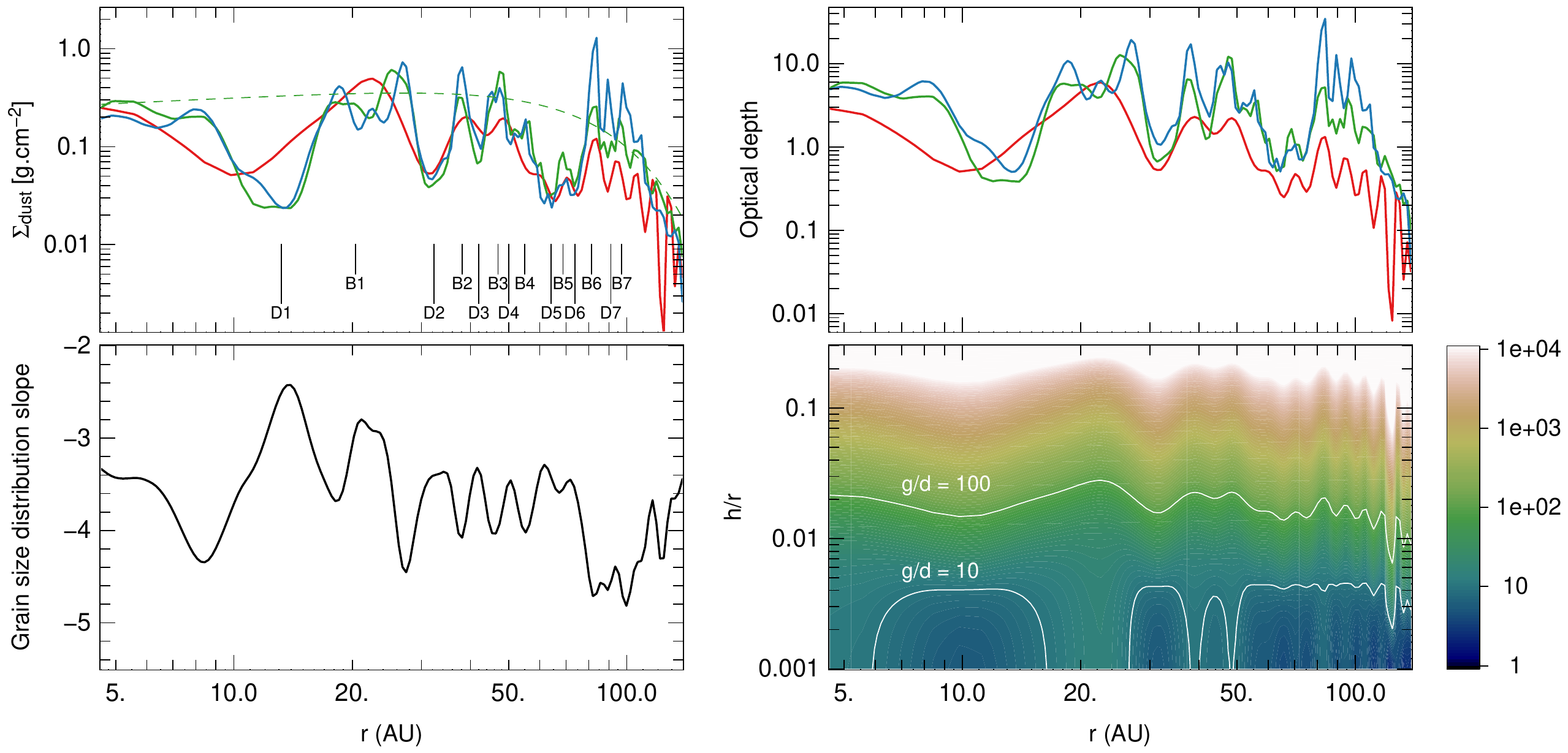}
    \caption{Structure of the disk model. In all panels, band 3 is shown as red, 6 as green, and 7 as blue lines. \emph{Top left:} dust surface density profiles obtained from inverting the brightness profiles along the disk major
    axis. The spatial resolution is 10, 4, and
    3\,AU respectively. The dashed line represents a tapered-edge model
      with $R_C = 80\,$AU and $\gamma = -0.2$ as derived by
      \cite{Kwon11,Kwon15} from CARMA 1.3\,mm data. The gaps and ring
      identified by \citetalias{ALMA_HLTau} are indicated.\emph{Top right:} vertical optical depth as a function of radius at the three
    wavelengths. \emph{Bottom left:} exponent of the grain size distribution
    within a column, using the combined data. The exponent is obtained by fitting a power-law to the
    actual local grain size distribution. \emph{Bottom right:} map of the
    gas-to-dust ratio. Only the disk is plotted here, not the atmosphere.
    \label{fig:Sigma} }
\end{figure*}

Figure~\ref{fig:Sigma} shows the resulting surface densities
obtained by inverting the brightness profiles of the best models for the different wavelengths.
The surface density profiles derived from observations in the three ALMA
bands are broadly consistent with each other, suggesting that our basic
modeling assumptions are also correct.
In particular, we find that the central peak as well as the two most central
rings are marginally optically thick, even at 2.9\,mm (Figure~\ref{fig:Sigma}),
while the outermost broad ring becomes optically thin. The surface density
extracted from the Band~6 data is also in good agreement  with the analytical
surface density derived by \cite{Kwon11} and \cite{Kwon15} from CARMA
observations at the same wavelength (green dashed line in
Figure~\ref{fig:Sigma}, in the bright rings,
the gaps were not detected by CARMA). In particular, our procedure recovers the transition
between the power-law description and exponential tapering at radii larger than
 $\approx\,80\,$AU, as well as the shape of surface density in both regimes.

Interestingly, assuming a grain size distribution slope of $p=-3.5$,
the surface density profiles we derive from the three bands start to
differ significantly  in the outermost ring, where the estimated surface density
decreases as a function of wavelength. A  model with a
steeper slope, $p=-4.5$,  results in similar surface density
profiles at all bands in the outer ring, but the densities now differ
significantly in the central regions.
This suggests that grains larger than a fraction of a millimeter are depleted in the outer ring
compared with the rest of the disk, supporting the idea  that grain growth is less
efficient in this part of the disk or that in the event of grain growth, radial
migration has already effectively removed a significant fraction of the large
grains. This can be observed directly by comparing the visibility curves
  at each wavelength (see bottom left panel of Figure~\ref{fig:obs_vs_mod}).
  The shorter the wavelength, the
  faster the visibilities approach 0 as the baseline increases, indicating that the emission is getting
  more and more compact as the wavelength increases.
Because the central rings are optically thick, this change in the
derived surface density as a function of wavelength could, at least
partly, also be due to the difference in optical depths.
However, we explored several models, changing the dust distribution,
but could not find a model that reproduces the surface density profile at all
wavelengths, without invoking a change of the
dust distribution as a function of the position.

To build a single model that reproduces the three bands simultaneously, we
adopt a procedure similar to the one used in \cite{Pinte07} and \cite{Gonzalez12}.
We use the three previously derived surface densities and assign each of them to a single grain
size, which emits most at the corresponding wavelength: the surface densities
from bands 3, 6, and 7 were used to define the distribution of the grains of size
0.4, 0.18, and 0.11\,mm respectively. For intermediate sizes, the spatial
distributions are interpolated in logarithm of the grain size. Outside of the
range 0.11-0.40\,mm, we do not extrapolate the distributions: all the grains
smaller than 0.11\,mm have the same surface density distribution, independently
of their size. Similarly, all the grains
larger than 0.40\,mm have the same surface density distribution.
Because a given grain size does not emit only in one given ALMA band,
  the final model does not show a as good agreement as the
individual models only fitting one wavelength. However, the emissivity as a function
of wavelength of a given
grain size remains peaked enough so that the contamination between bands remains
limited and the general agreement
is good at all wavelengths (Figure~\ref{fig:obs_vs_mod}).
The size distribution at any
point is no longer a power-law. In  bottom left panel of Figure~\ref{fig:Sigma}, we show the slope that
fits best the local (integrated vertically) grain size distribution. The grain
distribution has a local slope close to -3.5 up to 75\,AU, while it approaches
-4.5 between 75 and 120\,AU, supporting the idea of depletion of large grains.
The total dust mass of the best model is 5\,10$^{-4}$\,M$_\odot$.

Table~\ref{tab:rings} summarizes the properties of the various rings. The ring
masses range from 7 to 120\,M$_\oplus$.  We use the same nomenclature as in
\citetalias{ALMA_HLTau} to name the gaps and rings. The ring masses are
calculated by integrating the mass between two successive gaps, and the
temperatures are the mass averaged dust temperatures. Uncertainties on the gap
masses were computed by varying the gap depth, \emph{i.e.} by multiplying the
surface density in the gap by a constant without varying its shape. This procedure was repeated for
various values of the constant until the gap was infinitely deep (the surface
density virtually reaches 0 in the gap) and until it disappeared (the surface
density in the gap reaches the interpolation between the rings). For each of
these models, synthetic MCFOST images were calculated, visibilities
computed using the {\sc CASA} simulator and the total $\chi^2$ calculated.
Due to the computing cost of the procedure (dominated by the calculations of the
visibilities in {\sc CASA}),
a full Bayesian analysis taking into account
the potential correlations between the densities in the various gaps and rings
is out of reach with current computing facilities.
Instead, we explored the mass of each gap independently.
This approach allowed us to get an estimate of the uncertainties on the gaps (and
rings) masses.
In Table~\ref{tab:rings}, we report
the interval range where the $\chi^2$ remains smaller than 1.5 the minimum
$\chi^2$ value. We find that this threshold corresponds to visible differences
in the binned visibility curves as well as residuals above 3\,$\sigma$ in
the CLEAN maps.
Note that the central peak
and ring B1 are optically thick even at 3\,mm, so the corresponding mass
estimates are lower limits.

The dust opacity and spectral index for each ring are computed for the
 integrated dust population in each ring (Table~\ref{tab:rings}). The absolute
 values of the opacity result from our choice of dust properties and
 should be considered with an uncertainty of a factor a few.
The outer rings display smaller opacities at 1\,mm and  spectral index (from
$\approx$ 1 in the central rings to 1.3-1.4 in the two outer rings B6 and B7) due to the
steeper grain size distribution and relative lack of millimeter-sized dust
grains. Note that the change in opacity index between the inner and outer rings
is smaller  than the observed change in spectral index (bottom panel of
Figure~\ref{fig:profiles}), as the low observed spectral index in the central regions
of the disk results from a combination of the lower opacity index and high optical depth effects.

\capstartfalse
\begin{deluxetable}{l@{\hspace{-2mm}}ccccc}
\tabletypesize{\scriptsize}
\tablecolumns{6}
\tablewidth{\hsize}
\tablecaption{Ring properties. \label{tab:rings}}
\tablehead{
\vspace{-1.5mm}& &   \colhead{Mass}& \colhead{Optically} & \colhead{Dust Opaci-} & \\
\vspace{-1.5mm}&  \colhead{Dust}&   & & & \colhead{Dust} \\
\vspace{-1.5mm}&   &    \colhead{Average} &
\colhead{Thick\,\tablenotemark{c}}
& \colhead{ty\,\tablenotemark{d} (cm$^2$.g$^{-1}$ } & \\
\vspace{-1.5mm}Ring\tablenotemark{a}  & \colhead{Mass} & & \colhead{}& \colhead{}& \colhead{opacity}\\
\vspace{-1.5mm}&  & \colhead{T$_\mathrm{dust}$}  &  \colhead{at} &
\colhead{ of dust) at} &  \\
\vspace{-1.5mm}& \colhead{(M$_\oplus$)\,\tablenotemark{b}}&  & & & \colhead{Index\,\tablenotemark{d}}\\
& \colhead{}& \colhead{(K)}&  \colhead{$\lambda\!=\!3\,$mm} &
 \colhead{$\lambda\!=\!1\,$mm} & \colhead{}
}
\startdata
Peak & $>$2.3         & 630 & $\times$ &  6.0 &  0.9 \\
B1   & $>$47          & 48 & $\times$ &  5.7 & 1.0\\
B2   & $53_{-23}^{+16}$ & 36 & marginally&  6.6 &  1.0\\
B3   & $30_{-16}^{+7}$  & 32 & marginally&  6.9 & 1.1\\
B4   & $62_{-22}^{+20}$ & 25 & \nodata&  6.8 & 1.0\\
B5   & $7_{-1.5}^{+1.7}$ & 22 & \nodata& 7.1 & 0.8 \\
B6   & $101_{-17}^{+28}$ & 20 & marginally&  4.6 & 1.4 \\
B7   & $123_{-24}^{+34}$ & 16 &  \nodata&  4.9 &  1.3
\enddata
\tablenotetext{}{\hspace{-2.5mm} {\bf Notes.}}
\tablenotetext{a}{We use the same nomenclature as in
\citetalias{ALMA_HLTau} to name the rings}
\tablenotetext{b}{Ring masses are
calculated by integrating the surface density between two successive gaps. Uncertainties
correspond to the range of masses where $\chi^2 \leq 1.5 \times \min \chi^2$
(see the text)}
\tablenotetext{c}{A ring is considered optically thick if its vertical
   optical depth is larger than 3, and marginally optically thick if it is
   larger than 1}
\tablenotetext{d}{The dust opacity and opacity index
  are computed from the integrated dust distribution within a ring.}
\end{deluxetable}
\capstarttrue

So far in the paper, the iterations on the best model have been carried out in
the image plane. ALMA provides images with good fidelity; nevertheless, we also check in the $uv$ plane
that the final model visibilities
are consistent with the data. We compare the visibility functions of the
data sets in all three bands and in the direction in
$u$,$v$ space corresponding to the disk major and minor axes. To improve SNR, this includes all data within
$\pm10^\circ$ of these angles.
The lower panels in Figure~\ref{fig:obs_vs_mod}
illustrate these functions at each wavelength (left panel) and along the
minor and major disk axis (right panel), showing that there
is good agreement between the model and observed visibilities.
In Figure~\ref{fig:obs_vs_mod} (top panel), we compare the full ALMA image combining bands 6
and 7 with the model as ''observed'' by the {\sc CASA}
simulator, assuming the $uv$ coverage of the actual observations. Much of the faint non-axisymmetric
structure seen in the image is reproduced by this simulation, indicating that this is not inherent
to the source. Residuals (top right panel) are found to be $<3\sigma$ or $<$1\% of the local
flux/beam in this image.

\subsection{Gap brightness profiles}

The gaps show very similar profiles at bands 6 and 7 (Figure~\ref{fig:Sigma})
indicating that they are well resolved at these two wavelengths.
As a consequence, we used the profile derived from the combined band 6+7 image
(with the highest signal-to-noise), to measure the properties of each gap
(Table~\ref{tab:gaps}). We define the depth of a gap as the ratio of the
surface density in the gap and the surface density just outside of it
(averaged on both sides). The gap width is defined as the width at half-depth.
These values, together with the properties of the bright rings (Table~\ref{tab:rings}), can be used to set up hydrodynamical simulations aiming at better
understanding the origin of the gaps in the disk of HL~Tau, although this is outside the scope of the present paper.

The gaps, however, will be ''filled in'' if they are not fully spatially resolved, and so this table gives
only a lower limit to their depth. If we assume that the disk surface density was a smooth function
before the gaps formed, then an estimate of the 'missing mass' in the gaps can
be made. The gaps are consistent with a few tens of $M_{\oplus}$ of solids
missing (see Table~\ref{tab:gaps}), which is typical of the core size required
for runaway gas accretion, leading to gas giants. The uncertainties on the
mass of the gaps were estimated using a procedure similar to the one described
in Section~\ref{sec:sigma}. The density of each rings was varied independently
until the $\chi^2$ reached a value larger than 1.5 times the minimum $\chi^2$.

\capstartfalse
\begin{deluxetable}{l@{\hspace{-3mm}}ccccc}
\tabletypesize{\scriptsize}
\tablecolumns{6}
\tablewidth{\hsize}
\tablecaption{Gap properties.\label{tab:gaps}}
\tablehead{
\vspace{-1.5mm}&  &  & \colhead{Missing} &  & \colhead{Gap}\\
\vspace{-1.5mm} & \colhead{Position}& \colhead{Midplane}& & \colhead{Gap} &\\
\vspace{-1.5mm}Gap\tablenotemark{a}& &  & \colhead{Dust
  Mass\tablenotemark{b}}& &  \colhead{Width}\\
\vspace{-1.5mm}& (AU) & \colhead{T$_\mathrm{dust}$ [K]} & &
\colhead{Depth\tablenotemark{c}} & \\
& & & (M$_\oplus$)& &  \colhead{(AU)}
}
\startdata
D1 & 13.2 & 45 &  $7.2_{-0.9}^{+1.7}$ & 18 & 12\\
D2 & 32.3 & 36 & $22_{-4.2}^{+3.1}$ & 16 & 11\\
D3 & 42.0 & 29 & $20_{-11}^{+12}$ &  6.9 &  6.6\\
D4 & 50.0 & 24 &  $7_{-5.0}^{+2.4}$ &  3.8 &  4.5\\
D5 & 64.2 & 23 &  $8_{-4.7}^{+0.9}$ &  8.0 &  12\\
D6 & 73.7 & 23 & $27_{-17}^{+10}$ & 12 &  8.1\\
D7 & 91.0 & 21 & $36_{-20}^{+11}$ & 11 &  9.9\\
D5+D6\tablenotemark{d} & \nodata & \nodata & $140_{-28}^{+15}$ & 16 & 23
\enddata
\tablenotetext{}{\hspace{-2.5mm} {\bf Notes.}}
\tablenotetext{a}{We use the same nomenclature as in
\citetalias{ALMA_HLTau} to name the gaps}
\tablenotetext{b}{The missing mass is calculated by integrating the
  difference between the surface density in the gap and the interpolated
  surface density between the surrounding two rings. Uncertainties
correspond to the range of masses where $\chi^2 \leq 1.5 \times \min \chi^2$
(see the text)}
\tablenotetext{c}{The gap depth is the ratio of the density in the
surrounding rings (average of both sides) divided by the density at the minimum
of the gap.}
\tablenotetext{d}{The values of D5+D6 correspond to the assumption that the two gaps, D5 and D6, are a
large unique gap.}
\end{deluxetable}
\capstarttrue

\subsection{Degree of dust settling}

\begin{figure*}
  \includegraphics[width=\hsize]{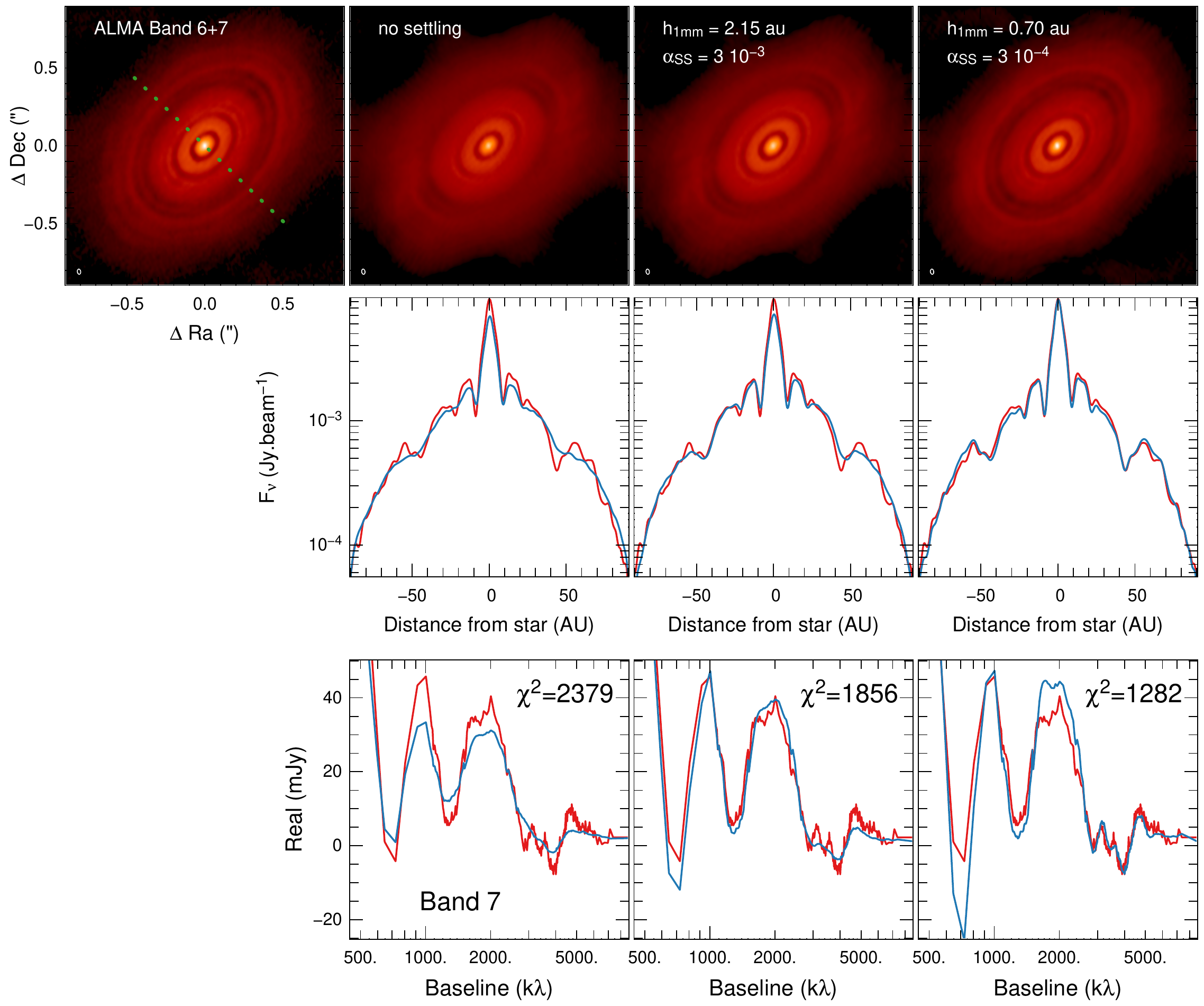}
  \caption{Effect of dust settling on the appearance of the images at
    1\,mm. This shows the enhanced contrast as the dust disk gets
    flatter. \emph{Upper row:} Band 6+7 CLEAN maps, \emph{left panel:} observations, and \emph{three panels on the right:}
    from left to right, model maps without settling, with
    $\alpha = 3\,10^{-3}$, and $3\,10^{-4}$. The corresponding
    scale height of the grains of 1\,mm in size is also indicated on each
    panel. \emph{Middle row:} cuts of the observed (red) and model (blue)
    CLEAN maps along the minor axis of the disk (green dashed line in top left
    panel). \emph{Bottom row:} real part of the Band 7 visibilities (first
    spectral window) along the disk
    minor axis, on an expanded scale (red lines are data, blue are the
    models). The visibilities at smaller baselines are not plotted here because
      they are practically not affected by dust settling. The indicated $\chi^2$
    were computed using the full data sets, \emph{i.e.} all the $u,v$ points in
    all the spectral windows of the three ALMA bands.
     \label{fig:settling}}
\end{figure*}

Dust settling in the disk of HL~Tau was suggested by \cite{Brittain05},
  based on the high gas-to-dust ratio in the upper layers of the disk, and by
  \cite{Kwon11} based on 1.3 and 2.6\,mm CARMA data and SED, but no
quantitative estimate was possible due to the limited spatial resolution.
Because the disk is inclined by $\sim$\,47$^\circ$ (with a variation of $\pm$\,1$^\circ$ between rings,
\citetalias{ALMA_HLTau}), the ALMA images enable us
to probe not only the radial but also the vertical extension of the various rings.  The fact
that they remain sharp and well-defined at all azimuthal angles
shows that the dust grains responsible for the emission are located in a thin layer in the midplane.
If the disk were thick, the gap contrast would be reduced - particularly
along the minor axis of the disk -- because of projection effects:
material in the nearby rings, located at significant heights above the midplane, would hide the gaps.
Figure~\ref{fig:settling} shows indeed that the appearance of the disk is strongly dependent on the degree of dust
settling. The variation of the width of the gaps as a function of azimuth is directly related to the thickness of
the emitting region (as well as to the beam shape, which is only slightly misaligned with the semi-minor axis).

A small value of the millimeter dust scale height, $h_{1\,\!\mathrm{mm}}$ = 0.70\,AU, equivalent to $\alpha = 3\times 10^{-4}$ produces
the best agreement with
most of the observed features in the ALMA image, with rings that are well
separated at all azimuthal angles. By contrast, in the absence of dust settling and for larger values of the
$\alpha$ parameter, \emph{e.g.}  $3\times10^{-3}$, dust of all sizes will
be better mixed with the gas in a geometrically thicker disk and the millimeter gap
contrast will be reduced.

The narrow ring (B5) inside the widest gap
(between the gaps D5 and D6, \citetalias{ALMA_HLTau}) shows
clear brightening at the ansae, due to its low optical depth.
However, the $h_{1\,\!\mathrm{mm}} = 0.70\,$AU model is not completely consistent with the ALMA observations: this ring seems
to disappear along the minor axis (this can also be seen in the difference between model and data in the
visibility plots along
the minor axis in Figure~\ref{fig:obs_vs_mod}). This is too large to be accounted for
by the difference in ring inclination angles. The model with $h_{1\,\!\mathrm{mm}}=2.15$\,AU reproduces
this behavior but does not provide a strong enough contrast in the other rings. This suggests that
the degree of vertical stirring of the dust grains may vary as a function of
the position tof he disk, either due to radial variation of  $\alpha$
or to local changes in the gas/dust ratio in the midplane (which were assumed
to be constant with radius here), or other stirring mechanism, like a big body in that gap.

\section{The gas disk: $^{12}$CO and HCO$^+$.}
\label{sec:gas}

\begin{figure*}
    \includegraphics[width=0.49\hsize]{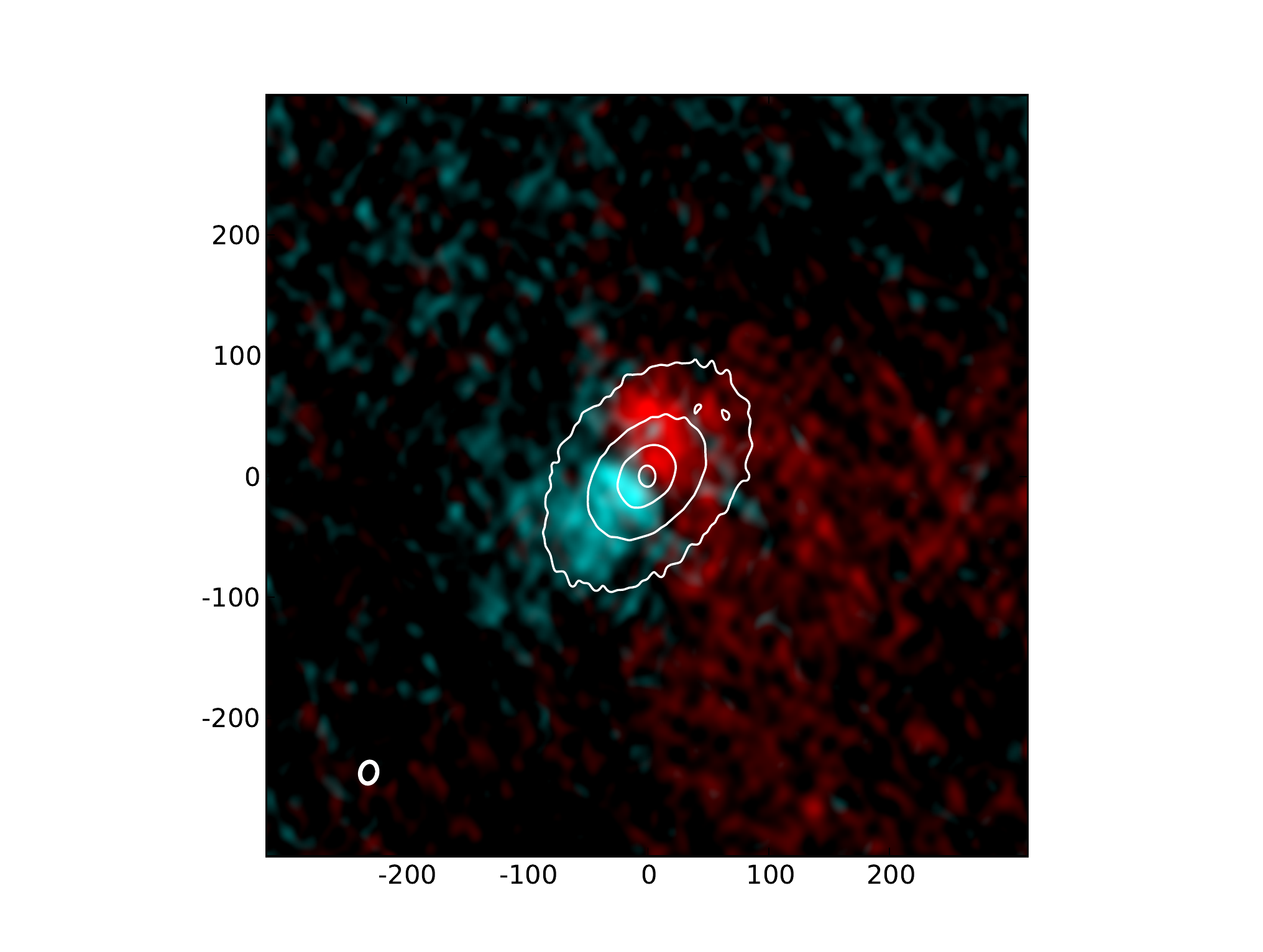}
    \includegraphics[width=0.49\hsize]{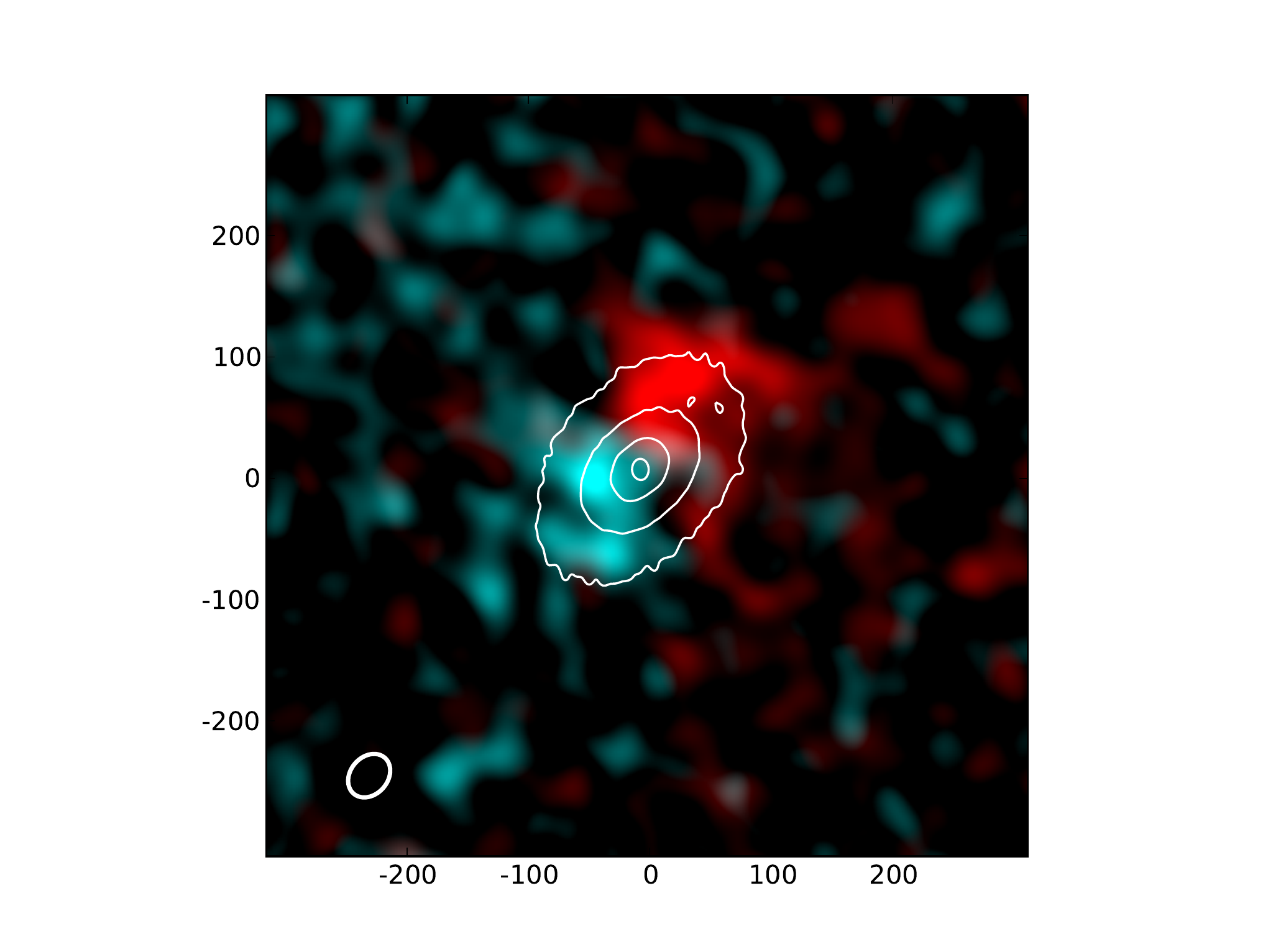}
    \caption{Distribution of red and
blue-shifted gas traced by (left) $^{12}$CO $J$=1-0 and (right) HCO$^+$ $J$=1-0, compared with
the dust continuum (shown by contours). The CO image shows integrated emission from -1 --
+5\,km.s$^{-1}$ and +10 -- +15km.s$^{-1}$ as cyan and red colors respectively. The HCO$^+$
image shows integrated emission from +2 -- +5\,km.s$^{-1}$ and +9 -- +13\,km.s$^{-1}$ as cyan
and red colors. The contours illustrate the continuum emission in band 3 at levels of
0.08, 0.24, 1.0, and 2.0\,mJy.beam$^{-1}$, which correspond approximately to
the outer edges of the major dust rings B7, B4, B1, and the bright central region. The
HCO$^+$ and continuum data sets are from the released ALMA science verification image
cubes, and the CO is from the ALMA visibility data, cleaned with a taper of 1600\,k$\lambda$
to give a resolution of $\sim$0.12\,arcsec.
Axes scale is given in AU and the resolutions of the molecular line
images are shown in the lower left corner of each panel. \label{fig:gas_im}}
\end{figure*}

HL Tau lies in a relatively dense part of the Taurus molecular cloud.
Emission from the extended ambient cloud contaminates the line profiles near
the cloud velocity of 6.8\,km.s$^{-1}$ and is resolved out on long
  baselines. However, as noted previously \citepalias{ALMA_HLTau}
a spatially compact component is seen in the ALMA images of
HCO$^+$ at velocities outside that of the narrow ambient emission, with a velocity
gradient consistent with disk rotation. Using the calibrated ALMA Science Verification data set
covering $^{12}$CO, we have employed {\sc CASA} CLEAN to make a spectral image with a
resolution of 0.12\,arcsec. Like HCO$^+$, this also shows a compact structure in the
high-velocity $^{12}$CO $J$=1-0 emission; Figure~\ref{fig:gas_im} compares the
integrated high-velocity blue- and red-shifted gas in the two molecules with the
dust emission contours. Both lines show relatively bright compact emission coincident with
the disk, with the highest velocity gradient toward the star and oriented
along the continuum major axis (orthogonal to the outflow direction). The CO image shows
faint extended red-shifted gas from the outflow to the southwest; however, most of the
compact high-velocity emission appears to be associated with the disk.

\begin{figure*}
\includegraphics[width=0.49\hsize]{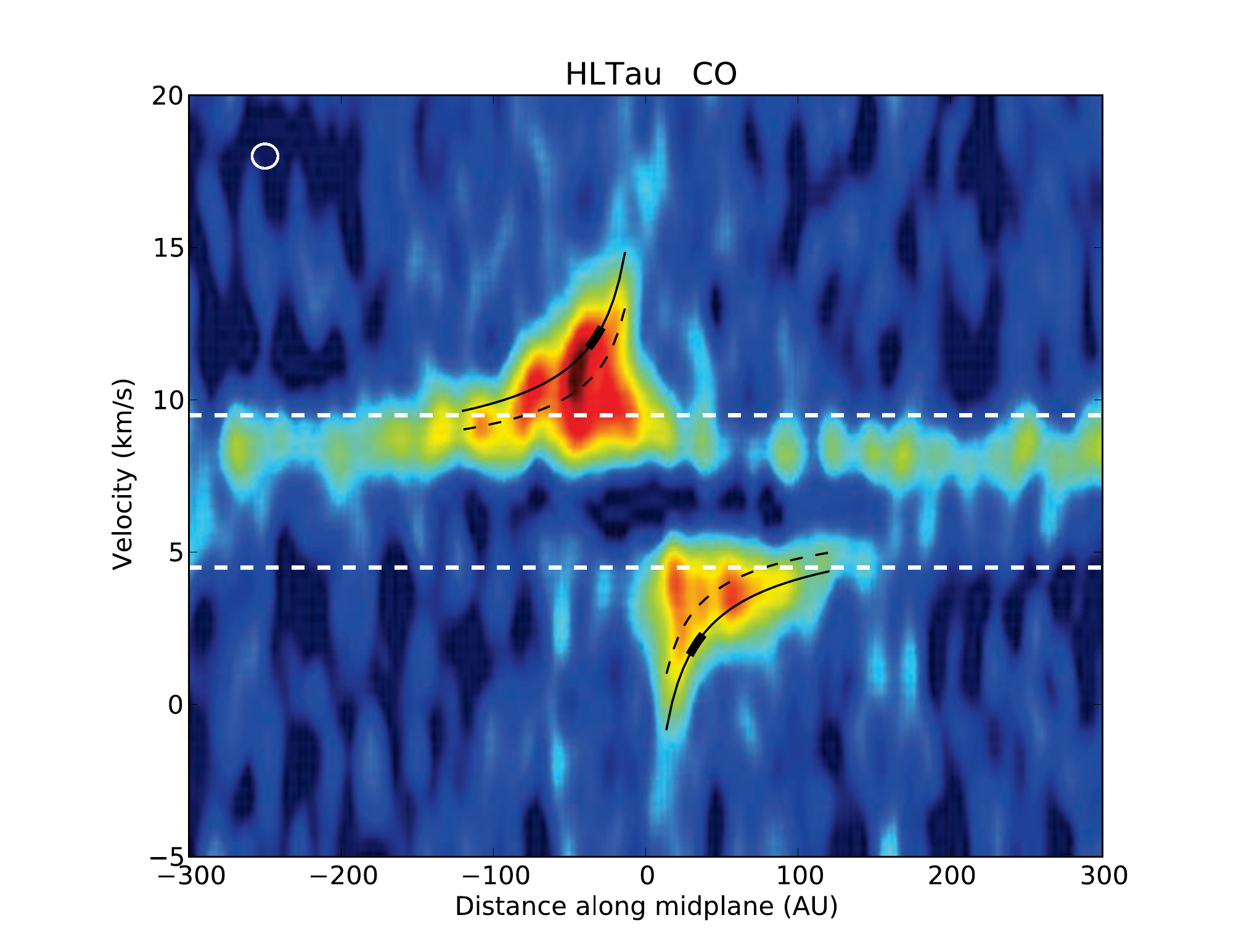}
\includegraphics[width=0.49\hsize]{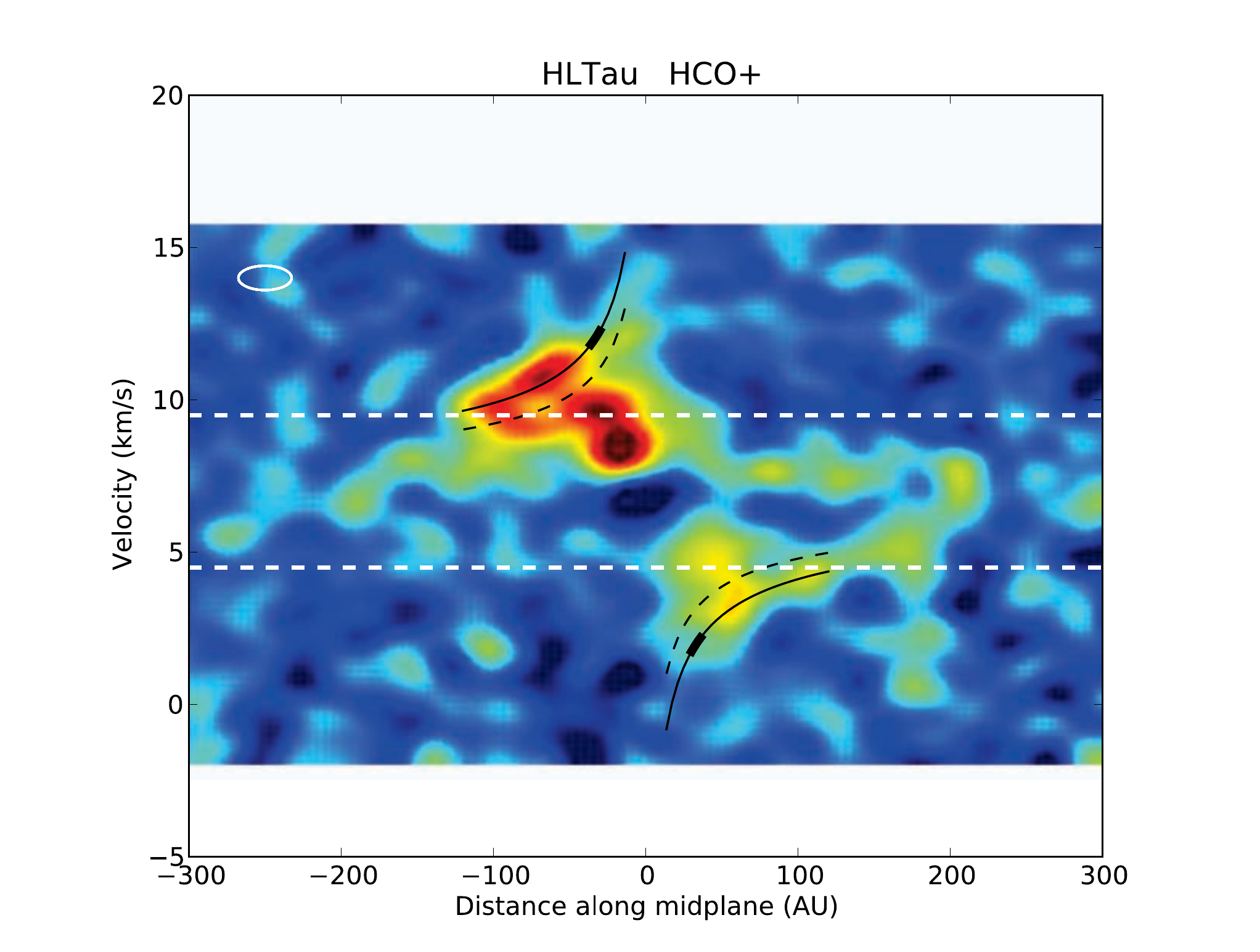}
\caption{Position--velocity diagrams of (left) $^{12}$CO $J$=1-0 and (right) HCO$^+$ $J$=1-0, taken along
the major axis of the disk. The spectra $\pm$ 0.15\,arcsec on either side of this major
axis have been averaged to form these PV diagrams.
The velocity extent where ambient cloud emission
is likely to contaminate the emission is indicated by the dashed white lines
(note that the ambient HCO$^+$ line is narrower,
so the contaminated velocity range is less than for $^{12}$CO).
The black lines show the Keplerian velocity for 1.7\,M$_{\odot}$ (solid)
and 1.0\,M$_{\odot}$ (dashed), with inner and outer radii of 10-120\,AU. On the 1.7\,M$_{\odot}$
line, the thick section represents
 the dark gap D2 in the millimeter dust disk (at radius of $\sim$\,35\,AU).
Spectral/spatial resolution
is shown upper left. The data have been spectrally smoothed to a resolution of
1\,km.s$^{-1}$.\label{fig:gas}}
\end{figure*}

In Figure~\ref{fig:gas}, we compare position-velocity diagrams along the disk major axis
in the two molecules,
with curves representing Keplerian rotation in a geometrically thin
disk. We use a disk inclination of 47$^\circ$, and show lines for enclosed masses of
1.7\,M\,$_{\odot}$ (solid) and 1.0\,M$_{\odot}$ (dashed).
Although not formal fits, the high gas velocities and shape of these PV
diagrams suggest that the stellar mass is closer
to $\sim$\,1.7\,M$_{\odot}$; this compares with
the estimates of 1.2\,M$_{\odot}$ based on HR tracks
\citep{Gudel07} and 1.3\,M$_{\odot}$ based on the approximate extent and width of
the HCO$^+$ emission line in \citetalias{ALMA_HLTau}.
The gas inner radius of 10\,AU is an upper limit set by the CO surface
brightness sensitivity at the highest velocities. The outer gas radius
is likely a lower limit because of the ambient cloud confusion at the low relative
velocities.
The location of the second
millimeter-grain gap (D2) is also indicated in the PV diagrams, and there is marginal evidence of an
increase in the CO and HCO$^+$ surface brightness beyond this radius.
Further studies and full
visibility modeling of the gas disk would be possible with a higher SNR data set at
similar resolutions to this data,  preferably using a molecular transition showing more contrast against
the ambient gas.

\section{Discussion - a coherent model for the HL Tau disk?}

We have produced a radiative transfer model that reproduces both the multi-wavelength
ALMA observations and SED of the
HL~Tau disk (see Appendix~\ref{app:SED}).  By inverting the
brightness maps obtained by ALMA, we can reconstruct the millimeter-grain surface density profile of
the disk. The central rings ($< 30$\,AU) appear marginally optically thick
even at 2.9\,mm, while the outermost ring becomes optically thin at the longest
wavelength.
The comparison of the three ALMA data sets between 870\,$\mu$mm and 3\,mm shows that this outer ring is also depleted in
the larger-sized grains compared to the more inner regions, suggesting that grain
growth is more efficient in the central regions or that some radial migration has already occurred
within the disk itself.

Because our model requires high optical depths at millimeter wavelengths,
  the calculated midplane temperatures in the
  gaps (Table~\ref{tab:gaps})
 are much lower than the temperatures used by \citet{Zhang15}, which
  assumed optically thin emission. As a result, in our model, the
 gaps location does
 not correspond to the condensation fronts of molecules.

To produce the observed gap/ring contrast, a decrease of the density of millimeter-sized
grains by a factor of $>$10 is required in the gaps compared to the adjacent
rings. The corresponding decrement in gas may be much more modest and detailed hydrodynamical
simulations of gas+dust systems are required to determine the mass of any
planet responsible for such gaps. Numerical predictions suggest that a
planetary core between 30\,M$_{\oplus}$ \citep{Paardekooper04} and 150\,M$_{\oplus}$ \citep{Fouchet10}
could produce such a decrement in the dust. This is similar to the 'missing
dust mass' in these gaps, as derived above.
The derived masses are also in agreement with the maximum planet mass
found by \cite{Tamayo15} in their nominal model:
40\,M$_{\oplus}$ if the planets are not in resonance, and 100\,M$_{\oplus}$
if the outer three planets are in a chain of 4:3 resonances. Dedicated HL~Tau
simulations by \cite{Dipierro15} indicate planet masses between $60$ and
$175$\,M$_{\oplus}$, which is, depending on the ring, of the same order or up to a
factor $\sim10$ larger than our mass estimate. The 'missing dust mass' may then
have formed the rocky cores of these planets.

The well-defined gaps observed by ALMA at all azimuthal angles clearly show
that the
millimeter-emitting dust grains are located in a geometrically thin
disk, with a scale height $h_{1\,\!\mathrm{mm}}$ at 100\,AU of no more than
2\,AU. This is  at least 5 times
smaller than the gas scale height assuming hydrostatic equilibrium (see the lower right panel in Figure~\ref{fig:Sigma}).
By using a simple settling model, we have reproduced the observations with a
turbulent viscosity coefficient $\alpha$ of $3 \times
10^{-4}$. Similar values have been derived by \cite{Mulders12} by fitting
  the median SEDs of three samples of protoplanetary disks surrounding Herbig
  stars, T Tauri stars, and brown dwarfs. For HL~Tau, we obtained this value by direct comparison with images.

The accretion rate of the disk surrounding HL~Tau was estimated to
  $8.7\,10^{-8}\,$M$_\odot$.yr$^{-1}$ from the Br$\gamma$ line luminosity
  \citep{Beck10}. Note that the Br$\gamma$ emission around HL~Tau is extended,
  so this value should be considered as an upper limit.
The viscosity coefficient can be estimated via
\begin{equation}
  \alpha(r) = \frac{\dot{M}}{3 \pi c_s(r)  h(r)  \Sigma(r)}
  \end{equation}
At $r=100\,$AU, we obtain $\alpha \approx 10^{-2}$, \emph{i.e.} at least one order of
magnitude larger than the value we estimated from the observed physical height of the
millimeter grains. Both estimates of $\alpha$ depend linearly on
the dust-to-gas mass ratio, which cannot account for the difference.

It is important to note that the two estimates of $\alpha$ do not measure the same
physical mechanism. The dust disk scale height depends on the diffusion coefficient,
\emph{i.e.} on the fluctuations of velocities and their correlation time \citep{Fromang06},
which are then converted to an $\alpha$ value assuming a Schmidt
number $S_C$, here set to 1.5, while the accretion rate gives an estimate of the transport
of angular momentum.
A larger  $S_C$ number, \emph{i.e.} a larger ratio of the momentum  and mass
diffusivities, can explain part of the disagreement.
Larger values than our assumed $S_C = 1.5$ have been reported from numerical
simulations e.g. 2.8 in \cite{Fromang06} and 11 in \cite{Carballido05}.
A simple $\alpha$-prescription for the
viscosity might also not be accurate enough to describe both mechanisms
simultaneously. In particular, the $\alpha$ coefficient is very likely not
constant throughout the disk and the estimation of $\dot{M}$ might be linked
to different parts of the disk. The dust diffusion process was also found to be
highly anisotropic, with larger values of $S_c$ in the radial direction than in
the vertical direction \citep[e.g.][]{Johansen06b}.

The SED of HL~Tau - particularly the far-infrared fluxes
and the observed extinction to the star -
cannot be reproduced with a hydrostatically supported disk.
Possible explanations include a large envelope, or alternatively, a puffed-up
disk atmosphere. The steep reddening gradient within 1\,arcsec of the disk noted by \cite{Close97}
suggests that most of the extinction is in the disk rather than an extended envelope.
As the gas and micron-sized grains are closely coupled, if
the gas scale height at 100\,AU was 30\,AU, it would be possible to fit the SED and get
such a large extinction gradient. This may be infalling gas, or part of
a photoevaporative or MHD wind \citep[e.g.][]{Gorti15}.

With the assumptions of our model, the gas-to-dust mass ratio in the disk midplane
is $\approx 6$ at a radius of 100\,AU, if we integrate
the dust size distribution up to 3\,mm. If  as expected,  grains have grown
further in the midplane, then the actual gas-to-dust ratio may be even
lower.
As an indication, and assuming our model can be extrapolated to larger sizes,
gas-to-dust ratio of 1 at 100\,AU is reached for $a_\mathrm{max} = 10\,$cm.

Because the millimeter dust disk is also very geometrically thin, these results provide direct constraints
on the models of solid growth in the disk midplane
\citep{Youdin02,Johansen06}. The increased dust density leads to a
vertical rotational velocity gradient near the midplane because of the feedback
(collision) of dust onto the gas. The
shear on the gas rotational velocity increases as the dust settles and the
gas/dust ratio decreases, leading eventually to
Kelvin-Helmholtz instabilities, which in turn will prevent the dust grains from settling further.
It has been suggested that, in this configuration,  streaming instabilities will lead to very
rapid clumping of dust particles
\citep{Youdin05,Johansen07b}, possibly
providing a mechanism to grow boulders rapidly.
The increased gas/dust ratio and flat dust sub-disk can also lead to "pebble
accretion'' \citep{Lambrechts12}, significantly reducing the growth timescale
of a planetary core massive enough to trigger runaway gas accretion.
In the calculations, turbulent diffusion and shear instabilities limit the amount of dust accumulation in the midplane, therefore
limiting the growth of the big bodies. Our results, although limited to
particle sizes of a few millimeters, give
specific indications regarding the size of the dust sub-disk and the decrease in the gas/dust ratio.

\section{Conclusions}

We have produced a model that reproduces the ALMA high-resolution observations
of HL Tau at the three wavelengths.
Detailed features including the rings, the gaps, their contrast and emissivity index can be modeled.
The results indicate that the gaps are devoid of dust by at least a factor of 10 compared to the adjacent rings.
The dust masses in the rings range from a few to $\sim 100\,M_{\oplus}$, and we estimate that if the disk
originally had a smooth radial distribution, up to 40$\,M_{\oplus}$ of dust has been removed
to create each of the two largest gaps.

Our modeling shows that the increased optical depth in the central rings
cannot explain alone the change in observed spectral index with radius. A change in the
dust properties is also required, which indicates that larger grains are
present closer to the star. This suggests that grain growth is faster in the
central regions and/or that radial migration is occurring in the disk. The
high optical depth of the disk in our model also means that the temperature in the gaps is
significantly lower than what was assumed by \cite{Zhang15} and that the gaps
do not seem to be coincident with molecular condensation fronts.

Interestingly, the sharp rings observed at all azimuthal angles clearly show that the disk emitting at
millimeter wavelengths is geometrically thin: of
order 1\,AU at 100\,AU.
We interpret this as a clear sign of dust vertical settling. Assuming a
  standard dust settling model, this implies that the coefficient of turbulent
  viscosity is of the order a few $10^{-4}$, between one and two orders of magnitude
  smaller than what is required to
  account for the estimated accretion on the protostar.
Because most
dust mass is in the larger grains, this would indicate that the gas/dust ratio
is about 5 in these rings.
For the first time (to our knowledge), we provide numbers obtained directly from
observations for the thickness and the decrease in the gas/dust
ratio in the disk mid-plane. By contrast, in order to fit the SED, the disk thickness in small grains and gas
may be significantly larger than hydrostatic equilibrium.

The parameters of the millimeter-grain disk are independent of the origins of the gaps and rings, but they will no doubt
provide significant constraints for planet formation models and/or HD or MHD models that will aim to reproduce them.

\begin{acknowledgements}
  This paper makes use of the following ALMA data: ADS/JAO.ALMA\#2011.0.00015.SV.
ALMA is a partnership of ESO (representing its member states), NSF (USA) and
NINS (Japan), together with NRC (Canada) and NSC and ASIAA (Taiwan),
in cooperation with the Republic of Chile. The Joint ALMA Observatory is
operated by ESO, AUI/NRAO, and NAOJ. The National Radio Astronomy Observatory is a facility of the
National Science Foundation operated under cooperative agreement by Associated
Universities, Inc. I.d.G. acknowledges support from MICINN (Spain)
AYA2011-30228-C03 grant (including FEDER funds). We thank C.~Dougados,
G.~Lesur, and H.~Klahr for useful discussions. The research leading to these
results has received funding from the European Union Seventh Framework
Programme FP7-2011 under grant agreement no 284405.
\end{acknowledgements}

\vspace{1cm}
\appendix
\section{Fitting simultaneously the SED and millimeter maps}
\label{app:SED}

\begin{figure}
  \includegraphics[width=\hsize]{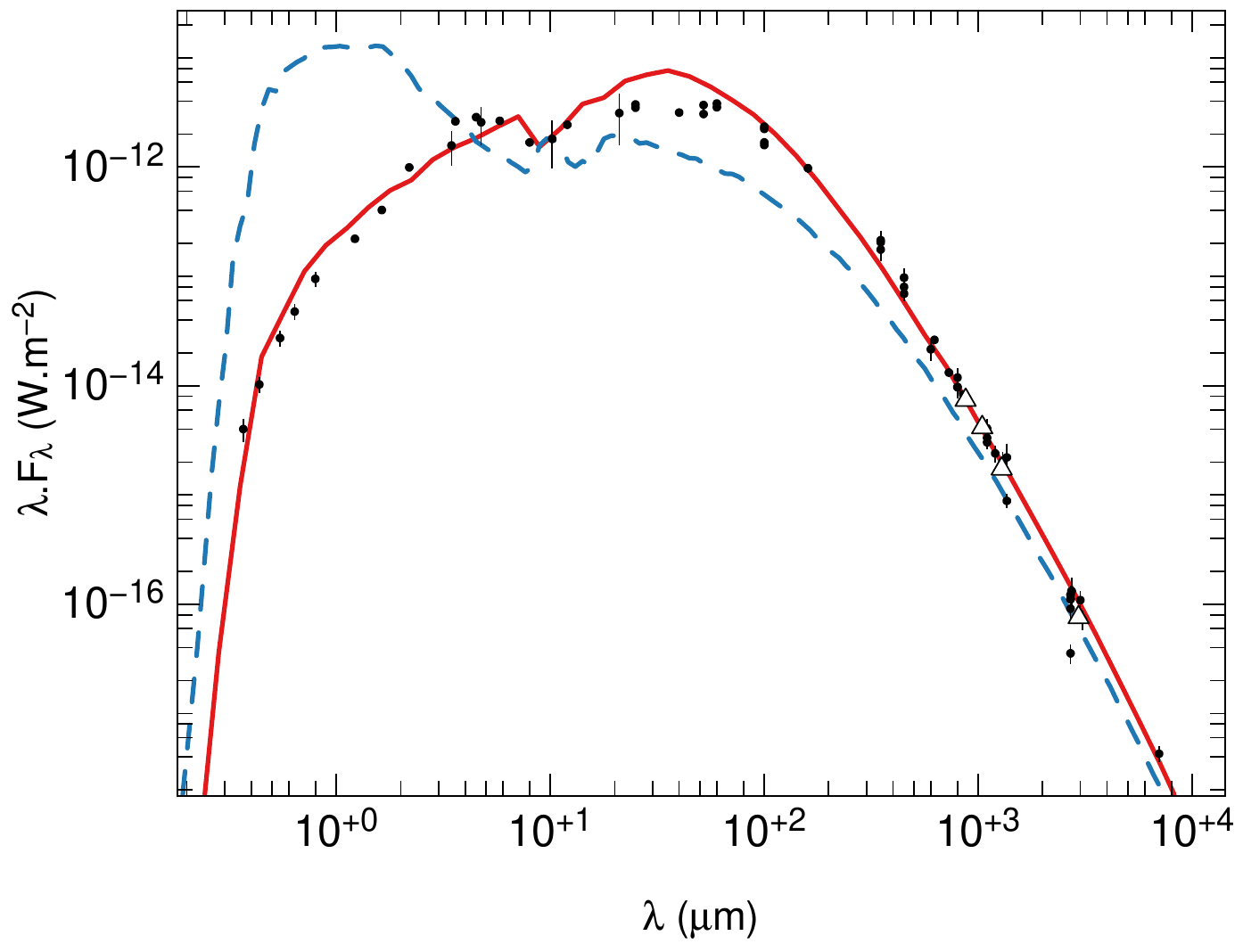}
  \caption{Spectral energy distribution of HL Tau. Photometry data points are
    from \cite{Robitaille07},  \cite{Menshchikov99}, and \citetalias{ALMA_HLTau}.
    The triangles show the fluxes obtained by ALMA. \emph{Red solid line:} disk + atmosphere,
    \emph{dashed blue line:} disk only.
\label{fig:SED}}
\end{figure}

Small (sub-micron) sized dust grains will be well-mixed with the gas. However, the simple model of a thin hydrostatic disk with a gas (and sub-micron dust) scale height $h_0\approx$10\,AU cannot reproduce the
SED, particularly in the far-infrared. Several authors have noted that an
additional component is needed to fit HL Tau, and generally this has been assumed to be a spherical
infalling envelope \citep{Menshchikov99,Robitaille07,Kwon11}. However, the ALMA
data \citepalias{ALMA_HLTau} also show that this system has
a broad CO bipolar outflow with an opening angle of $\sim80^\circ$ - indicating
that such an envelope would not be a spherical structure. The inclination of
the system suggests that we are looking along a line of sight somewhere between
this outflow cone and the $h_0$ = 10\,AU gas scale height proposed above. Several high-density
molecular tracers (CN, HCN, and HCO$^+$) show narrow line-of-sight
absorption against the bright disk continuum, and the line profiles suggest the gas may be blue-shifted by $\sim\,0.5$\,km.s$^{-1}$ \citepalias{ALMA_HLTau}. This would imply that the gas
high up in the disk atmosphere is actually outflowing rather than infalling.
The line-of-sight gas column
to the star has also been probed by infrared CO lines \citep{Brittain05}, and a
comparison of
the derived gas and dust column density was used to show that this material is dust-{\it depleted},
\emph{i.e.} has a gas/dust ratio of $>$100.
Because the photometry from the optical to mid-infrared regime suffers
  from high optical depth extinction, it is not possible to assess the exact
  nature of the dust structure responsible for this extinction. Several dust spatial
  distributions can account for the optical and near-infrared extinction and
  mid-infrared excess above the disk emission.
We chose to add a thick disk atmosphere composed of interstellar-like dust grains
(same composition as in the disk but with $a_\mathrm{max} = 1\,\mu$m). We find a
good agreement with the observed SED for
$h_0 = 30$\,AU, a surface density slope of -0.75 and dust mass of 2.8\,10$^{-6}$\,M$_\odot$. The
resulting model SED is presented in Figure~\ref{fig:SED}.
It does not
require the addition of a spherical envelope, but does require a disk
in the sub-micron sized grains that is considerably thicker than hydrostatic equilibrium.
Because of its low mass, compared to the disk, and small opacity at millimeter wavelength, this
  extra component is undetectable in the synthetic ALMA maps and does not affect the
  results presented in this paper.

\section{Hydrostatic scale height}
\label{app:h_hydro}

\begin{figure}
  \includegraphics[width=\hsize]{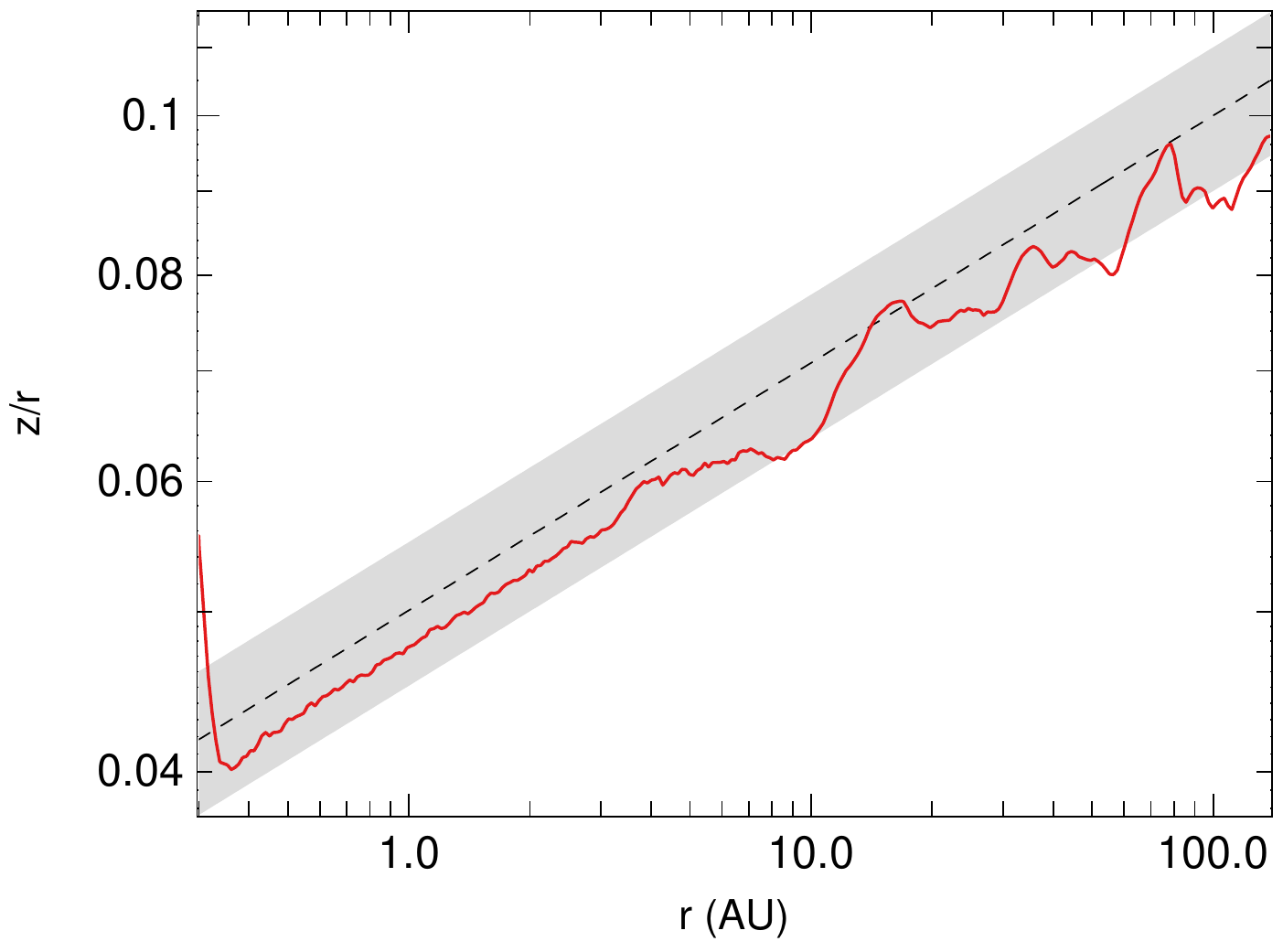}
  \caption{Gas vertical scale height. The full red  line
represents the hydrostatic scale height computed from the midplane temperature
of the best model. The dashed black line represents the gas scale height used
as an input for the model, \emph{i.e.} 10\,AU at 100\,AU.
The shaded area represents variations of the scale height of 10\,\%,
\emph{i.e.} between 9 and 11\,AU at 100\,AU.
\label{fig:eq_hydro}}
\end{figure}

In the modeling presented in this paper, we have assumed a fixed gas vertical
structure. Figure~\ref{fig:eq_hydro} presents
the calculated midplane temperatures compared to the temperature
corresponding to the gas scale height of the best model.
The agreement is very good almost everywhere in the disk, except at the very
inner edge where stellar radiation heats the midplane directly.
In the outer disk, the actual disk temperature oscillates due to the alternating
rings and gaps, but the variation remains limited and corresponds to a change
in the scale height of about 5\,\%.

\bibliographystyle{yahapj.bst}
\bibliography{biblio}

\begin{thebibliography}{}
\providecommand\natexlab[1]{#1}
\providecommand\JournalTitle[1]{#1}

\bibitem[{{ALMA Partnership} {et~al.}(2015{\natexlab{a}}){ALMA Partnership},
  {Fomalont}, {Vlahakis}, {Corder}, {Remijan}, {Barkats}, {Lucas}, {Hunter},
  {Brogan}, \& {Asaki Y.~et~al.}}]{ALMA_LBC}
{ALMA Partnership}, {Fomalont}, E.~B., {Vlahakis}, C., {et~al.}
  2015{\natexlab{a}},
  \href{http://dx.doi.org/10.1088/2041-8205/808/1/L1}{\JournalTitle{\apjl},
  808, L1}

\bibitem[{{ALMA Partnership} {et~al.}(2015{\natexlab{b}}){ALMA Partnership},
  {Brogan}, {P{\'e}rez}, {Hunter}, {Dent}, {Hales}, {Hills}, {Corder},
  {Fomalont}, {Vlahakis}, \& {Asaki Y.~et~al.}}]{ALMA_HLTau}
{ALMA Partnership}, {Brogan}, C.~L., {P{\'e}rez}, L.~M., {et~al.}
  2015{\natexlab{b}},
  \href{http://dx.doi.org/10.1088/2041-8205/808/1/L3}{\JournalTitle{\apjl},
  808, L3}

\bibitem[{{Andrews} {et~al.}(2009){Andrews}, {Wilner}, {Hughes}, {Qi}, \&
  {Dullemond}}]{Andrews09}
{Andrews}, S.~M., {Wilner}, D.~J., {Hughes}, A.~M., {Qi}, C., \& {Dullemond},
  C.~P. 2009,
  \href{http://dx.doi.org/10.1088/0004-637X/700/2/1502}{\JournalTitle{\apj},
  700, 1502}

\bibitem[{{Barri{\`e}re-Fouchet} {et~al.}(2005){Barri{\`e}re-Fouchet},
  {Gonzalez}, {Murray}, {Humble}, \& {Maddison}}]{Barriere05}
{Barri{\`e}re-Fouchet}, L., {Gonzalez}, J.-F., {Murray}, J.~R., {Humble},
  R.~J., \& {Maddison}, S.~T. 2005,
  \href{http://dx.doi.org/10.1051/0004-6361:20042249}{\JournalTitle{\aap}, 443,
  185}

\bibitem[{{Beck} {et~al.}(2010){Beck}, {Bary}, \& {McGregor}}]{Beck10}
{Beck}, T.~L., {Bary}, J.~S., \& {McGregor}, P.~J. 2010,
  \href{http://dx.doi.org/10.1088/0004-637X/722/2/1360}{\JournalTitle{\apj},
  722, 1360}

\bibitem[{{Beckwith} {et~al.}(1990){Beckwith}, {Sargent}, {Chini}, \&
  {Guesten}}]{Beckwith90}
{Beckwith}, S.~V.~W., {Sargent}, A.~I., {Chini}, R.~S., \& {Guesten}, R. 1990,
  \href{http://dx.doi.org/10.1086/115385}{\JournalTitle{\aj}, 99, 924}

\bibitem[{{Birnstiel} {et~al.}(2010){Birnstiel}, {Ricci}, {Trotta},
  {Dullemond}, {Natta}, {Testi}, {Dominik}, {Henning}, {Ormel}, \&
  {Zsom}}]{Birnstiel10}
{Birnstiel}, T., {Ricci}, L., {Trotta}, F., {et~al.} 2010,
  \href{http://dx.doi.org/10.1051/0004-6361/201014893}{\JournalTitle{\aap},
  516, L14}

\bibitem[{{Brittain} {et~al.}(2005){Brittain}, {Rettig}, {Simon}, \&
  {Kulesa}}]{Brittain05}
{Brittain}, S.~D., {Rettig}, T.~W., {Simon}, T., \& {Kulesa}, C. 2005,
  \href{http://dx.doi.org/10.1086/429310}{\JournalTitle{\apj}, 626, 283}

\bibitem[{{Carballido} {et~al.}(2005){Carballido}, {Stone}, \&
  {Pringle}}]{Carballido05}
{Carballido}, A., {Stone}, J.~M., \& {Pringle}, J.~E. 2005,
  \href{http://dx.doi.org/10.1111/j.1365-2966.2005.08850.x}{\JournalTitle{\mnras},
  358, 1055}

\bibitem[{{Chandler} \& {Richer}(2000)}]{Chandler00}
{Chandler}, C.~J., \& {Richer}, J.~S. 2000,
  \href{http://dx.doi.org/10.1086/308401}{\JournalTitle{\apj}, 530, 851}

\bibitem[{{Close} {et~al.}(1997){Close}, {Roddier}, {J.~Northcott}, {Roddier},
  \& {Elon Graves}}]{Close97}
{Close}, L.~M., {Roddier}, F., {J.~Northcott}, M., {Roddier}, C., \& {Elon
  Graves}, J. 1997, \JournalTitle{\apj}, 478, 766

\bibitem[{{D'Alessio} {et~al.}(1998){D'Alessio}, {Canto}, {Calvet}, \&
  {Lizano}}]{DAlessio98}
{D'Alessio}, P., {Canto}, J., {Calvet}, N., \& {Lizano}, S. 1998,
  \href{http://dx.doi.org/10.1086/305702}{\JournalTitle{\apj}, 500, 411}

\bibitem[{{Dipierro} {et~al.}(2015){Dipierro}, {Price}, {Laibe}, {Hirsh},
  {Cerioli}, \& {Lodato}}]{Dipierro15}
{Dipierro}, G., {Price}, D., {Laibe}, G., {et~al.} 2015,
  \href{http://dx.doi.org/10.1093/mnrasl/slv105}{\JournalTitle{\mnras}, 453,
  L73}

\bibitem[{{Dominik} {et~al.}(2007){Dominik}, {Blum}, {Cuzzi}, \&
  {Wurm}}]{Dominik07PPV}
{Dominik}, C., {Blum}, J., {Cuzzi}, J.~N., \& {Wurm}, G. 2007,
  \JournalTitle{Protostars and Planets V}, 783

\bibitem[{{Dorschner} {et~al.}(1995){Dorschner}, {Begemann}, {Henning},
  {Jaeger}, \& {Mutschke}}]{Dorschner95}
{Dorschner}, J., {Begemann}, B., {Henning}, T., {Jaeger}, C., \& {Mutschke}, H.
  1995, \JournalTitle{\aap}, 300, 503

\bibitem[{{Duch{\^e}ne} {et~al.}(2010){Duch{\^e}ne}, {McCabe}, {Pinte},
  {Stapelfeldt}, {M{\'e}nard}, {Duvert}, {Ghez}, {Maness}, {Bouy},
  {Navascu{\'e}s}, {Morales-Calder{\'o}n}, {Wolf}, {Padgett}, {Brooke}, \&
  {Noriega-Crespo}}]{Duchene10}
{Duch{\^e}ne}, G., {McCabe}, C., {Pinte}, C., {et~al.} 2010,
  \href{http://dx.doi.org/10.1088/0004-637X/712/1/112}{\JournalTitle{\apj},
  712, 112}

\bibitem[{{Dullemond} \& {Dominik}(2005)}]{dd05}
{Dullemond}, C.~P., \& {Dominik}, C. 2005,
  \href{http://dx.doi.org/10.1051/0004-6361:20042080}{\JournalTitle{\aap}, 434,
  971}

\bibitem[{{Dullemond} {et~al.}(2002){Dullemond}, {van Zadelhoff}, \&
  {Natta}}]{Dullemond02}
{Dullemond}, C.~P., {van Zadelhoff}, G.~J., \& {Natta}, A. 2002,
  \href{http://dx.doi.org/10.1051/0004-6361:20020608}{\JournalTitle{\aap}, 389,
  464}

\bibitem[{{Fouchet} {et~al.}(2010){Fouchet}, {Gonzalez}, \&
  {Maddison}}]{Fouchet10}
{Fouchet}, L., {Gonzalez}, J.-F., \& {Maddison}, S.~T. 2010,
  \href{http://dx.doi.org/10.1051/0004-6361/200913778}{\JournalTitle{\aap},
  518, A16}

\bibitem[{{Fouchet} {et~al.}(2007){Fouchet}, {Maddison}, {Gonzalez}, \&
  {Murray}}]{Fouchet07}
{Fouchet}, L., {Maddison}, S.~T., {Gonzalez}, J.-F., \& {Murray}, J.~R. 2007,
  \href{http://dx.doi.org/10.1051/0004-6361:20077586}{\JournalTitle{\aap}, 474,
  1037}

\bibitem[{{Fromang} \& {Nelson}(2009)}]{Fromang09}
{Fromang}, S., \& {Nelson}, R.~P. 2009,
  \href{http://dx.doi.org/10.1051/0004-6361/200811220}{\JournalTitle{\aap},
  496, 597}

\bibitem[{{Fromang} \& {Papaloizou}(2006)}]{Fromang06}
{Fromang}, S., \& {Papaloizou}, J. 2006,
  \href{http://dx.doi.org/10.1051/0004-6361:20054612}{\JournalTitle{\aap}, 452,
  751}

\bibitem[{{Garaud} {et~al.}(2004){Garaud}, {Barri{\`e}re-Fouchet}, \&
  {Lin}}]{Garaud2004}
{Garaud}, P., {Barri{\`e}re-Fouchet}, L., \& {Lin}, D.~N.~C. 2004,
  \href{http://dx.doi.org/10.1086/381385}{\JournalTitle{\apj}, 603, 292}

\bibitem[{{Gonzalez} {et~al.}(2012){Gonzalez}, {Pinte}, {Maddison},
  {M{\'e}nard}, \& {Fouchet}}]{Gonzalez12}
{Gonzalez}, J.-F., {Pinte}, C., {Maddison}, S.~T., {M{\'e}nard}, F., \&
  {Fouchet}, L. 2012,
  \href{http://dx.doi.org/10.1051/0004-6361/201218806}{\JournalTitle{\aap},
  547, A58}

\bibitem[{{Gorti} {et~al.}(2015){Gorti}, {Hollenbach}, \&
  {Dullemond}}]{Gorti15}
{Gorti}, U., {Hollenbach}, D., \& {Dullemond}, C.~P. 2015,
  \href{http://dx.doi.org/10.1088/0004-637X/804/1/29}{\JournalTitle{\apj}, 804,
  29}

\bibitem[{{Greaves} {et~al.}(2008){Greaves}, {Richards}, {Rice}, \&
  {Muxlow}}]{Greaves08}
{Greaves}, J.~S., {Richards}, A.~M.~S., {Rice}, W.~K.~M., \& {Muxlow}, T.~W.~B.
  2008,
  \href{http://dx.doi.org/10.1111/j.1745-3933.2008.00559.x}{\JournalTitle{\mnras},
  391, L74}

\bibitem[{{G{\"u}del} {et~al.}(2007){G{\"u}del}, {Briggs}, {Arzner}, {Audard},
  {Bouvier}, {Feigelson}, {Franciosini}, {Glauser}, {Grosso}, {Micela},
  {Monin}, {Montmerle}, {Padgett}, {Palla}, {Pillitteri}, {Rebull}, {Scelsi},
  {Silva}, {Skinner}, {Stelzer}, \& {Telleschi}}]{Gudel07}
{G{\"u}del}, M., {Briggs}, K.~R., {Arzner}, K., {et~al.} 2007,
  \href{http://dx.doi.org/10.1051/0004-6361:20065724}{\JournalTitle{\aap}, 468,
  353}

\bibitem[{{Guilloteau} {et~al.}(2011){Guilloteau}, {Dutrey}, {Pi{\'e}tu}, \&
  {Boehler}}]{Guilloteau11}
{Guilloteau}, S., {Dutrey}, A., {Pi{\'e}tu}, V., \& {Boehler}, Y. 2011,
  \href{http://dx.doi.org/10.1051/0004-6361/201015209}{\JournalTitle{\aap},
  529, A105}

\bibitem[{{Johansen} {et~al.}(2006{\natexlab{a}}){Johansen}, {Henning}, \&
  {Klahr}}]{Johansen06}
{Johansen}, A., {Henning}, T., \& {Klahr}, H. 2006{\natexlab{a}},
  \href{http://dx.doi.org/10.1086/502968}{\JournalTitle{\apj}, 643, 1219}

\bibitem[{{Johansen} {et~al.}(2006{\natexlab{b}}){Johansen}, {Klahr}, \&
  {Mee}}]{Johansen06b}
{Johansen}, A., {Klahr}, H., \& {Mee}, A.~J. 2006{\natexlab{b}},
  \href{http://dx.doi.org/10.1111/j.1745-3933.2006.00191.x}{\JournalTitle{\mnras},
  370, L71}

\bibitem[{{Johansen} \& {Youdin}(2007)}]{Johansen07b}
{Johansen}, A., \& {Youdin}, A. 2007,
  \href{http://dx.doi.org/10.1086/516730}{\JournalTitle{\apj}, 662, 627}

\bibitem[{{Kwon} {et~al.}(2011){Kwon}, {Looney}, \& {Mundy}}]{Kwon11}
{Kwon}, W., {Looney}, L.~W., \& {Mundy}, L.~G. 2011,
  \href{http://dx.doi.org/10.1088/0004-637X/741/1/3}{\JournalTitle{\apj}, 741,
  3}

\bibitem[{{Kwon} {et~al.}(2015){Kwon}, {Looney}, {Mundy}, \& {Welch}}]{Kwon15}
{Kwon}, W., {Looney}, L.~W., {Mundy}, L.~G., \& {Welch}, W.~J. 2015,
  \href{http://dx.doi.org/10.1088/0004-637X/808/1/102}{\JournalTitle{\apj},
  808, 102}

\bibitem[{{Lambrechts} \& {Johansen}(2012)}]{Lambrechts12}
{Lambrechts}, M., \& {Johansen}, A. 2012,
  \href{http://dx.doi.org/10.1051/0004-6361/201219127}{\JournalTitle{\aap},
  544, A32}

\bibitem[{{Lay} {et~al.}(1997){Lay}, {Carlstrom}, \& {Hills}}]{Lay97}
{Lay}, O.~P., {Carlstrom}, J.~E., \& {Hills}, R.~E. 1997,
  \href{http://dx.doi.org/10.1086/304815}{\JournalTitle{\apj}, 489, 917}

\bibitem[{{Looney} {et~al.}(2000){Looney}, {Mundy}, \& {Welch}}]{Looney00}
{Looney}, L.~W., {Mundy}, L.~G., \& {Welch}, W.~J. 2000,
  \href{http://dx.doi.org/10.1086/308239}{\JournalTitle{\apj}, 529, 477}

\bibitem[{{Men'shchikov} {et~al.}(1999){Men'shchikov}, {Henning}, \&
  {Fischer}}]{Menshchikov99}
{Men'shchikov}, A.~B., {Henning}, T., \& {Fischer}, O. 1999,
  \href{http://dx.doi.org/10.1086/307333}{\JournalTitle{\apj}, 519, 257}

\bibitem[{{Menu} {et~al.}(2014){Menu}, {van Boekel}, {Henning}, {Chandler},
  {Linz}, {Benisty}, {Lacour}, {Min}, {Waelkens}, {Andrews}, {Calvet},
  {Carpenter}, {Corder}, {Deller}, {Greaves}, {Harris}, {Isella}, {Kwon},
  {Lazio}, {Le Bouquin}, {M{\'e}nard}, {Mundy}, {P{\'e}rez}, {Ricci},
  {Sargent}, {Storm}, {Testi}, \& {Wilner}}]{Menu14}
{Menu}, J., {van Boekel}, R., {Henning}, T., {et~al.} 2014,
  \href{http://dx.doi.org/10.1051/0004-6361/201322961}{\JournalTitle{\aap},
  564, A93}

\bibitem[{{Mulders} \& {Dominik}(2012)}]{Mulders12}
{Mulders}, G.~D., \& {Dominik}, C. 2012,
  \href{http://dx.doi.org/10.1051/0004-6361/201118127}{\JournalTitle{\aap},
  539, A9}

\bibitem[{{Mundt} {et~al.}(1990){Mundt}, {Buehrke}, {Solf}, {Ray}, \&
  {Raga}}]{Mundt90}
{Mundt}, R., {Buehrke}, T., {Solf}, J., {Ray}, T.~P., \& {Raga}, A.~C. 1990,
  \JournalTitle{\aap}, 232, 37

\bibitem[{{Mundy} {et~al.}(1996){Mundy}, {Looney}, {Erickson}, {Grossman},
  {Welch}, {Forster}, {Wright}, {Plambeck}, {Lugten}, \& {Thornton}}]{Mundy96}
{Mundy}, L.~G., {Looney}, L.~W., {Erickson}, W., {et~al.} 1996,
  \href{http://dx.doi.org/10.1086/310117}{\JournalTitle{\apjl}, 464, L169}

\bibitem[{{Paardekooper} \& {Mellema}(2004)}]{Paardekooper04}
{Paardekooper}, S.-J., \& {Mellema}, G. 2004,
  \href{http://dx.doi.org/10.1051/0004-6361:200400053}{\JournalTitle{\aap},
  425, L9}

\bibitem[{{Pagani} {et~al.}(2010){Pagani}, {Steinacker}, {Bacmann}, {Stutz}, \&
  {Henning}}]{Pagani10}
{Pagani}, L., {Steinacker}, J., {Bacmann}, A., {Stutz}, A., \& {Henning}, T.
  2010,
  \href{http://dx.doi.org/10.1126/science.1193211}{\JournalTitle{Science}, 329,
  1622}

\bibitem[{{P{\'e}rez} {et~al.}(2012){P{\'e}rez}, {Carpenter}, {Chandler},
  {Isella}, {Andrews}, {Ricci}, {Calvet}, {Corder}, {Deller}, {Dullemond},
  {Greaves}, {Harris}, {Henning}, {Kwon}, {Lazio}, {Linz}, {Mundy}, {Sargent},
  {Storm}, {Testi}, \& {Wilner}}]{Perez12}
{P{\'e}rez}, L.~M., {Carpenter}, J.~M., {Chandler}, C.~J., {et~al.} 2012,
  \href{http://dx.doi.org/10.1088/2041-8205/760/1/L17}{\JournalTitle{\apjl},
  760, L17}

\bibitem[{{Pinte} {et~al.}(2007){Pinte}, {Fouchet}, {M{\'e}nard}, {Gonzalez},
  \& {Duch{\^e}ne}}]{Pinte07}
{Pinte}, C., {Fouchet}, L., {M{\'e}nard}, F., {Gonzalez}, J.-F., \&
  {Duch{\^e}ne}, G. 2007,
  \href{http://dx.doi.org/10.1051/0004-6361:20077137}{\JournalTitle{\aap}, 469,
  963}

\bibitem[{{Pinte} {et~al.}(2009){Pinte}, {Harries}, {Min}, {Watson},
  {Dullemond}, {Woitke}, {M{\'e}nard}, \& {Dur{\'a}n-Rojas}}]{Pinte09}
{Pinte}, C., {Harries}, T.~J., {Min}, M., {et~al.} 2009,
  \href{http://dx.doi.org/10.1051/0004-6361/200811555}{\JournalTitle{\aap},
  498, 967}

\bibitem[{{Pinte} \& {Laibe}(2014)}]{Pinte14}
{Pinte}, C., \& {Laibe}, G. 2014,
  \href{http://dx.doi.org/10.1051/0004-6361/201220545}{\JournalTitle{\aap},
  565, A129}

\bibitem[{{Pinte} {et~al.}(2008{\natexlab{a}}){Pinte}, {M{\'e}nard}, {Berger},
  {Benisty}, \& {Malbet}}]{Pinte08}
{Pinte}, C., {M{\'e}nard}, F., {Berger}, J.~P., {Benisty}, M., \& {Malbet}, F.
  2008{\natexlab{a}},
  \href{http://dx.doi.org/10.1086/527378}{\JournalTitle{\apjl}, 673, L63}

\bibitem[{{Pinte} {et~al.}(2006){Pinte}, {M{\'e}nard}, {Duch{\^e}ne}, \&
  {Bastien}}]{Pinte06}
{Pinte}, C., {M{\'e}nard}, F., {Duch{\^e}ne}, G., \& {Bastien}, P. 2006,
  \href{http://dx.doi.org/10.1051/0004-6361:20053275}{\JournalTitle{\aap}, 459,
  797}

\bibitem[{{Pinte} {et~al.}(2008{\natexlab{b}}){Pinte}, {Padgett}, {M{\'e}nard},
  {Stapelfeldt}, {Schneider}, {Olofsson}, {Pani{\'c}}, {Augereau},
  {Duch{\^e}ne}, {Krist}, {Pontoppidan}, {Perrin}, {Grady}, {Kessler-Silacci},
  {van Dishoeck}, {Lommen}, {Silverstone}, {Hines}, {Wolf}, {Blake}, {Henning},
  \& {Stecklum}}]{Pinte08b}
{Pinte}, C., {Padgett}, D.~L., {M{\'e}nard}, F., {et~al.} 2008{\natexlab{b}},
  \href{http://dx.doi.org/10.1051/0004-6361:200810121}{\JournalTitle{\aap},
  489, 633}

\bibitem[{{Ricci} {et~al.}(2010){Ricci}, {Testi}, {Natta}, {Neri}, {Cabrit}, \&
  {Herczeg}}]{Ricci10}
{Ricci}, L., {Testi}, L., {Natta}, A., {et~al.} 2010,
  \href{http://dx.doi.org/10.1051/0004-6361/200913403}{\JournalTitle{\aap},
  512, A15}

\bibitem[{{Robitaille} {et~al.}(2007){Robitaille}, {Whitney}, {Indebetouw}, \&
  {Wood}}]{Robitaille07}
{Robitaille}, T.~P., {Whitney}, B.~A., {Indebetouw}, R., \& {Wood}, K. 2007,
  \href{http://dx.doi.org/10.1086/512039}{\JournalTitle{\apjs}, 169, 328}

\bibitem[{{Shakura} \& {Sunyaev}(1973)}]{ShakuraSunyaev1973}
{Shakura}, N.~I., \& {Sunyaev}, R.~A. 1973, \JournalTitle{\aap}, 24, 337

\bibitem[{{Stephens} {et~al.}(2014){Stephens}, {Looney}, {Kwon},
  {Fern{\'a}ndez-L{\'o}pez}, {Hughes}, {Mundy}, {Crutcher}, {Li}, \&
  {Rao}}]{Stephens14}
{Stephens}, I.~W., {Looney}, L.~W., {Kwon}, W., {et~al.} 2014,
  \href{http://dx.doi.org/10.1038/nature13850}{\JournalTitle{\nat}, 514, 597}

\bibitem[{{Tamayo} {et~al.}(2015){Tamayo}, {Triaud}, {Menou}, \&
  {Rein}}]{Tamayo15}
{Tamayo}, D., {Triaud}, A.~H.~M.~J., {Menou}, K., \& {Rein}, H. 2015,
  \href{http://dx.doi.org/10.1088/0004-637X/805/2/100}{\JournalTitle{\apj},
  805, 100}

\bibitem[{{Testi} {et~al.}(2014){Testi}, {Birnstiel}, {Ricci}, {Andrews},
  {Blum}, {Carpenter}, {Dominik}, {Isella}, {Natta}, {Williams}, \&
  {Wilner}}]{Testi14PPVI}
{Testi}, L., {Birnstiel}, T., {Ricci}, L., {et~al.} 2014,
  \href{http://dx.doi.org/10.2458/azu_uapress_9780816531240-ch015}{\JournalTitle{Protostars
  and Planets VI}, 339}

\bibitem[{{Wilner} {et~al.}(1996){Wilner}, {Ho}, \& {Rodriguez}}]{Wilner96}
{Wilner}, D.~J., {Ho}, P.~T.~P., \& {Rodriguez}, L.~F. 1996,
  \href{http://dx.doi.org/10.1086/310307}{\JournalTitle{\apjl}, 470, L117}

\bibitem[{{Youdin} \& {Goodman}(2005)}]{Youdin05}
{Youdin}, A.~N., \& {Goodman}, J. 2005,
  \href{http://dx.doi.org/10.1086/426895}{\JournalTitle{\apj}, 620, 459}

\bibitem[{{Youdin} \& {Shu}(2002)}]{Youdin02}
{Youdin}, A.~N., \& {Shu}, F.~H. 2002,
  \href{http://dx.doi.org/10.1086/343109}{\JournalTitle{\apj}, 580, 494}

\bibitem[{{Zhang} {et~al.}(2015){Zhang}, {Blake}, \& {Bergin}}]{Zhang15}
{Zhang}, K., {Blake}, G.~A., \& {Bergin}, E.~A. 2015,
  \href{http://dx.doi.org/10.1088/2041-8205/806/1/L7}{\JournalTitle{\apjl},
  806, L7}

\bibitem[{{Zubko} {et~al.}(1996){Zubko}, {Mennella}, {Colangeli}, \&
  {Bussoletti}}]{Zubko96}
{Zubko}, V.~G., {Mennella}, V., {Colangeli}, L., \& {Bussoletti}, E. 1996,
  \JournalTitle{\mnras}, 282, 1321

\end{thebibliography}

\end{document}